\documentclass[iop]{emulateapj}
\pdfoutput=1
\usepackage{amsmath}
\usepackage{graphics,graphicx}
\usepackage{natbib}
\usepackage{multirow}
\usepackage[percent]{overpic}
\usepackage{color}
\usepackage{chemformula}

\newcounter{chem}
\newcounter{temp}

\newenvironment{chequation}{%
  \setcounter{temp}{\value{equation}}%
  \setcounter{equation}{\value{chem}}%
  \renewcommand{\theequation}{R\arabic{equation}}%
}{%
  \setcounter{chem}{\value{equation}}%
  \setcounter{equation}{\value{temp}}%
}
\begin{document}

%% Short title and author list
\shorttitle{Pale Orange Dots}
\shortauthors{Arney et al.}

\title{Pale Orange Dots: The Impact of Organic Haze on the Habitability and Detectability of Earthlike Exoplanets}

%% \author and \affil call.
\author{Giada N. Arney\altaffilmark{1,2,3,4,5}, Victoria S. Meadows\altaffilmark{1,2,3}, Shawn D. Domagal-Goldman\altaffilmark{2,4}, Drake Deming\altaffilmark{2,6}, Tyler D. Robinson\altaffilmark{2,7}, Guadalupe Tovar\altaffilmark{1}, Eric T. Wolf\altaffilmark{8}, Edward Schwieterman\altaffilmark{1,2,3,5,9,10}}

\altaffiltext{1}{University of Washington Astronomy Department, Box 351580, U.W. Seattle, WA 98195; giada.n.arney@nasa.gov}
\altaffiltext{2}{NASA Astrobiology Institute Virtual Planetary Laboratory, Box 351580, U.W. Seattle, WA 98195}
\altaffiltext{3}{University of Washington Astrobiology Program, Box 351580, U.W. Seattle, WA 98195}
\altaffiltext{4}{NASA Goddard Space Flight Center, 8800 Greenbelt Road, Greenbelt, MD 20771}
\altaffiltext{5}{NASA Postdoctoral Program, Universities Space Research Association, Columbia, Maryland, USA.}
\altaffiltext{6}{University of Maryland Department of Astronomy, College Park, MD 20742}
\altaffiltext{7}{Department of Astronomy and Astrophysics, University of California, Santa Cruz, CA 95064}
\altaffiltext{8}{University of Colorado at Boulder Laboratory for Astrophysics and Space Physics, 1234 Innovation Drive, Boulder, CO 80303}
\altaffiltext{9}{University of California at Riverside, Riverside, California, USA.}
\altaffiltext{10}{Blue Marble Space Institute of Science 1200 Westlake Ave N Suite 1006, Seattle, WA 98109}

\begin{abstract}
Hazes are common in known planet atmospheres, and geochemical evidence suggests early Earth occasionally supported an organic haze with significant environmental and spectral consequences. The UV spectrum of the parent star drives organic haze formation through methane photochemistry. We use a 1D photochemical-climate model to examine production of fractal organic haze on Archean Earth-analogs in the habitable zonesof several stellar types: the modern and early Sun, AD Leo (M3.5V), GJ 876 (M4V), $\epsilon$ Eridani (K2V), and $\sigma$ Bo{\"o}tis (F2V). For Archean-like atmospheres, planets orbiting stars with the highest UV fluxes do not form haze due to the formation of photochemical oxygen radicals that destroy haze precursors. Organic hazes impact planetary habitability via UV shielding and surface cooling, but this cooling is minimized around M dwarfs whose energy is emitted at wavelengths where organic hazes are relatively transparent. We generate spectra to test the detectability of haze. For 10 transits of a planet orbiting GJ 876 observed by the James Webb Space Telescope, haze makes gaseous absorption features at wavelengths $<$ 2.5 $\mu$m 2-10$\sigma$ shallower compared to a haze-free planet, and methane and carbon dioxide are detectable at $>$5$\sigma$. A haze absorption feature can be detected at 5$\sigma$ near 6.3 $\mu$m, but higher signal-to-noise is needed to distinguish haze from adjacent absorbers. For direct imaging of a planet at 10 parsecs using a coronagraphic 10-meter class ultraviolet-visible-near infrared telescope, a UV-blue haze absorption feature would be strongly detectable at $>$12$\sigma$ in 200 hours. 
\end{abstract}


\section{Introduction}
We stand at the brink of a revolution in comparative planetology, with observations of potentially habitable terrestrial planets possible within the next decades. We have discovered far more exoplanets, and of different types, than the worlds in our solar system. Statistics derived from the Kepler sample \citep[e.g.][]{Borucki2010} suggest it is likely that a non-transiting and transiting Earth-sized planet (1-1.5 R$_{Earth}$) orbits in the habitable zones of M dwarf stars within 2.6 and 10.6 parsecs (pc), respectively \citep{Dressing2015}, and an Earth-sized planet has been discovered in the habitable zone of Proxima Centauri \citep{Anglada2016}. Another estimate puts the fraction of potentially habitable Earth-sized plants orbiting M dwarfs as high as 0.8 per star \citep{Morton2014}. The Transiting Exoplanet Survey Satellite (TESS) will search for these nearby worlds, and a handful of them may be observable with upcoming space-based missions such as the James Webb Space Telescope (JWST) and the Wide-Field Infrared Survey Telescope (WFIRST) \citep{Beichman2014, Spergel2015}. In the coming decades, dedicated large space telescope concepts currently under consideration including the Large UV-Optical-IR Surveyor (LUVOIR) and the Habitable Exoplanet Imaging Mission (HabEx) may allow us to directly image a larger sample of potentially habitable worlds, explore their chemical diversities, and search for biosignatures in their reflected light spectra \citep{Postman2010, Bolcar2015, Dalcanton2015, Seager2015, Stapelfeldt2015, Mennesson2016}. 

Attempts to characterize exoplanets with future telescopes may be frustrated by the presence of atmospheric hazes, so it is important to understand which planet and star combinations are more likely to form them. We have observed evidence for hazes or clouds in the transit transmission spectra of several exoplanets \citep{Bean2010, Sing2011, Knutson2014, Knutson2014a, Kreidberg2014, DeWit2016}. In fact, the only sub-Neptune-sized planet currently known to have an obviously clear atmosphere is HAT-P-11b \citep{Fraine2014}. However, the planets that have been characterized thus far are sub-Neptunes or larger and orbit close to their host stars. From observations of planets in our own solar system we know that photochemical hazes, whose formation is driven by UV radiation from the sun, occur frequently in the atmospheres of small worlds: Venus has a thick deck of H$_2$SO$_4$ cloud and haze, Titan is completely obscured by orange organic hazes, and even Pluto has thin yet multi-layered organic hazes \citep{Rannou2009}. 

The best constrained example of a hazy terrestrial planet in the habitable zone of its parent star is provided by the ancient Earth. During the Archean eon (3.8-2.5 billion years ago), an intermittently-present organic haze similar to Titan's may have existed in our planet's atmosphere \citep{Pavlovkerogen2001, Trainer2004, Trainer2006, DeWitt2009, Hasenkopf2010, Zerkle2012, Kurzweil2013, Claire2014, Izon2015, Hicks2016}. Geochemical evidence for this haze centers around correlations between sulfur and organic carbon isotopes which imply that the surface UV flux was attenuated while the atmospheric redox state remained reducing. A UV-absorbing organic haze straightforwardly explains these environmental conditions. The Archean Earth can serve as an archetype for organic-rich hazy, habitable exoplanets. 

Methane-rich terrestrial exoplanets with organic haze may occur frequently as methane (CH$_4$) can be produced by several abiotic processes \citep{Kasting2005, Kelley2005, Etiope2013, Guzman2013}, and CH$_4$-producing metabolisms (i.e., methanogenesis) are simple and evolved early on Earth \citep{Woese1977, Kharecha2005, Ueno2006}.  In Archean Earth's atmosphere, organic haze could have formed when the ratio of methane to carbon dioxide (CO$_2$) in the atmosphere was above 0.1, and its formation under Archean-analog atmospheric conditions has been observed in the laboratory \citep{DeWitt2009, Trainer2006, Hasenkopf2010, Hasenkopf2011, Hicks2016}. 

The interactions of haze with incoming sunlight would have had important climatic consequences for our early planet, and this consideration will be relevant to hazy exoplanets \citep{Pavlov2001, Domagal-Goldman2008, Haqq-Misra2008, Wolf2010, Hasenkopf2011, Arney2016}. Organic haze, whose formation is initiated by methane photochemistry, would have scattered and absorbed incoming solar radiation, heating the stratosphere while cooling the planet's surface \citep{McKay1991}. The cooling effects associated with geologically-constrained CO$_2$ abundances and an Archean haze under a fainter young sun \citep{Sagan1997} might suggest surface conditions too cold to support life, but our previous climate modeling work on the hazy Archean does not support this. We used paleosol constraints on CO$_2$ measured by \citet{Driese2011} for the partial pressure of CO$_2$ (pCO$_2$) in the Archean atmosphere at 2.7 billion years ago (Ga), pCO$_2$ = 0.0036-0.018 bars. Despite our conservative CO$_2$ estimate, we previously found that habitable conditions are possible at 2.7 Ga for 0.5 bar and 1 bar atmospheres, even with the fainter young Sun \citep{Arney2016}. We found that habitable conditions are possible under a haze for three reasons: firstly, we used fractal-shaped (Section \ref{models}) rather than spherical particles, which result in less cooling \citep{Wolf2010}; secondly, haze formation was found to be self-limiting due to UV self-shielding which shuts off haze formation; thirdly, we revised our lower temperature limit for habitability based on the results of 3D climate modeling studies which show that planets like Archean Earth can maintain stable open ocean belts at global average temperatures below freezing \citep{Wolf2013} and as low as 250 K \citep{Charnay2013}.

Our previous simulations for Archean haze production were for Earth orbiting the 2.7 Ga Sun. Here, we expand on this earlier work to test organic haze formation under the influence of other stellar spectra using the same self-consistent photochemical-climate simulations we employed in our previous work \citep{Arney2016}. We present an analysis of organic haze production for Archean-analog planets orbiting several types of stars. To help guide future telescope observations of hazy habitable exoplanets, we use instrument simulators with realistic noise sources for JWST and a future 10-meter class space telescope (LUVOIR) to predict the detectability of spectral features from haze-rich atmospheres.

\section{Models and Methods}\label{models}
To simulate Archean-analog planets orbiting various types of stars, we use a coupled 1D photochemical-climate model called $\mathtt{Atmos}$ to simulate photochemical hazes and examine their climatic effects. Hazy spectra are generated using our 1D line-by-line fully multiple scattering radiative transfer model, $\mathtt{SMART}$, \citep[the Spectral Mapping and Atmosphere Radiative Transfer Model,][]{Meadows1996, Crisp1997}.  $\mathtt{SMART}$ has been validated against multiple solar system planets \citep[e.g.,][]{Robinson2011, Arney2014}. Synthetic spectra with realistic noise estimates for JWST and a large aperture coronagraph telescope are generated using the models described in \citet{Deming2009} and \citet{Robinson2016}. $\mathtt{Atmos}$ and $\mathtt{SMART}$ are described in detail in \citet{Arney2016}, but we provide a summary of them here, beginning with a description of our haze treatment. 

\begin{figure*}[!htb]
\begin{center}
\includegraphics [scale=1]{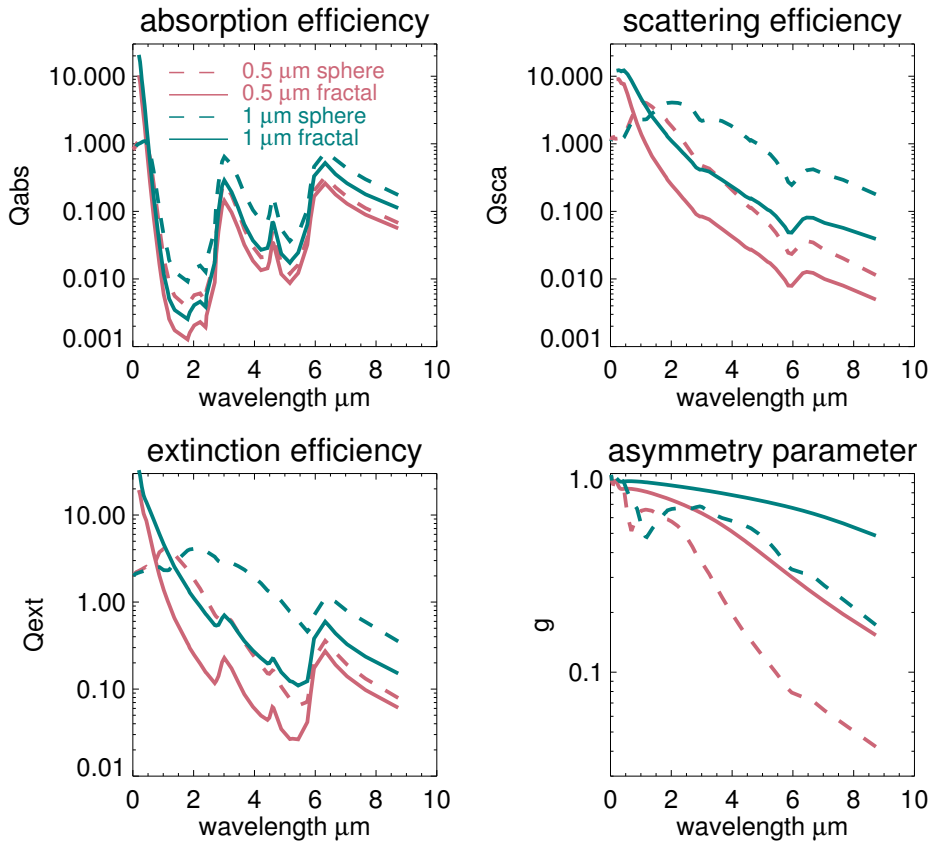}\\
\caption{The absorption efficiency (Q$_{abs}$), scattering efficiency (Q$_{sca}$), extinction efficiency (Q$_{ext}$ = Q$_{abs}$ + Q$_{sca}$), and the asymmetry parameter (g) as a function of wavelength for 0.5 $\mu$m (pink) and 1 $\mu$m (teal) fractal (solid lines) versus spherical (dashed lines) particles. $g$ is a measure of the degree of particle forward scattering; a value of 1 would indicate perfect forward-scattering, while 0 would indicate perfectly isotropic scattering. Fractal particles produce more extinction at short wavelengths than equal mass spherical particles, and they also tend to be more forward scattering. This diminishes their ability to cool the planet by allowing longer wavelengths to be transmitted to the surface. We assume optical constants from \citet{Khare1984}.}
\label{fig:pod2_1}
\end{center}
\end{figure*}

Our models simulate haze particles as fractal, rather than spherical, in shape. Studies of Titan's atmosphere suggest that fractal particles are a more realistic shape for organic hazes compared to spherical Mie particles \citep{Rannou1997}. Fractal particles are composed of multiple smaller spherical ``monomers'' clumped together into a larger aggregate, and their scattering and absorbing behavior differs from spherical particles. Short wavelengths interact with the small monomers, while longer wavelengths interact with the bulk aggregate, and the net result is that fractal particles produce more extinction at ultraviolet (UV) wavelengths and less at visible and infrared (IR) wavelengths compared to equal mass spherical particles. For the particle scattering physics in this work, we adopt the fractal mean-field approximation \citep{Botet1997} based on the work of \citet{Wolf2010}. The mean field approximation has been validated through studies of silica aggregates \citep{Botet1997} and the haze in Titan's atmosphere \citep{Rannou1997, Larson2015}. Figure \ref{fig:pod2_1} shows a comparison of the wavelength-dependent optical properties of spherical versus fractal particles for two different particle sizes assuming optical constants from \citet{Khare1984}.The fractal particles shown in this figure have masses equal to spherical particles of radius 0.5 $\mu$m and 1 $\mu$m. Fractal particles tend to produce more forward-scattering and have less overall extinction than equal-mass spherical particles -- except at the shortest wavelengths. 

The climate portion of the $\mathtt{Atmos}$ model was originally developed by \citet{Kasting1986}, but has evolved considerably since this first version.  The version we use in this study was most recently used to re-calculate habitable zone boundaries around main sequence stars \citep{Kopparapu2013} and to study the climatic consequences of hazes in the Archean atmosphere \citep{Arney2016}.  In this latter study, the scattering and absorption properties of fractal hazes were incorporated into the model to augment its existing spherical (Mie) haze capabilities.  

The photochemical code is based on one developed by \citet{Kasting1979} but significantly modernized by \citet{Zahnle2006}. The model can use different stellar spectra as inputs. Flux from wavelengths spanning 8 angstroms wide on either side of Lyman alpha (121.6 nm) is binned into the model ``Lyman alpha'' bin. This photochemical model was recently modified to include fractal hydrocarbon hazes \citep{Zerkle2012}. A complete list of chemical reactions and species boundary conditions for our Archean model is in the supplementary materials of \citet{Arney2016}. The photochemical model's aerosol formation scheme follows the method described in \citet{Pavlov2001}. Because the full chemical pathways to haze formation are not yet understood \citep[e.g.,][]{Hallquist2009}, the model uses a simplified chemical scheme to form haze particles. In this scheme, C$_2$H + C$_2$H$_2$ $\rightarrow$ C$_4$H$_2$ + H and C$_2$H + CH$_2$CCH$_2$ $\rightarrow$ C$_5$H$_4$ + H lead directly to haze particle formation, with the C$_4$H$_2$ and C$_5$H$_4$ immediately condensing out as particles. 

Our haze formation scheme has limitations which could cause the haze formation rate to be over- or under-predicted. This scheme follows a mechanism proposed for the formation of Titan's hazes \citep{Allen1980, Yung1984, Pavlov2001} and does not include the incorporation of oxygen or nitrogen atoms into haze particles expected for early Earth's hazes. Experiments generating hazes using ultraviolet radiation (115-400 nm) have shown that haze formation occurring in a CH$_4$/CO$_2$/N$_2$ mixture can exceed the haze formation rate in a pure CH$_4$/N$_2$ mixture (Trainer et al. 2006), which could lead to our model under-predicting the haze formation rate since we do not include oxygen incorporation into haze molecules. Another study has shown that oxygen derived from CO$_2$ can constitute 10\% of the mass of Archean-analog haze particles \citep{Hicks2016}. Our model also does not include the ion chemistry known to be important to the formation of Titan's hazes, which may also lead to an under-estimation of haze formation \citep[e.g.][]{Waite2007}. On the other hand, C$_4$H$_2$ can revert back to C$_2$H$_2$ in a real atmosphere, but this is not included in our photochemical scheme since C$_4$H$_2$ is assumed to condense out as haze particles, and this could lead to our model over-estimating haze-production. Our future work will include model updates that will allow us to include these important effects.

Outputs from $\mathtt{Atmos}$ are used as inputs to our radiative transfer code, $\mathtt{SMART}$ \citep{Meadows1996, Crisp1997} to produce synthetic spectra. The haze is included in $\mathtt{SMART}$ via a particle size binning scheme described in \citet{Arney2016}. To generate transit transmission spectra, we use the $\mathtt{SMART-T}$ version of the model \citep{Misra2014, Misra2014a}, which includes the refraction effects, geometry, and correspondingly longer path lengths of transit observations. 

To simulate JWST observations, we use the model described in \citet{Deming2009}, and our simulated observations employ the JWST parameters described in \citet{Schwieterman2016}. The JWST model we use can simulate the Mid-Infrared Instrument \citep[MIRI,][]{Wright2004}, the Near-Infrared Spectrograph \citep[NIRSpec,][]{Ferruit2012} and the Near Infrared Imager and Slitless Spectrograph \citep[NIRISS,][]{Doyon2012}. The simulator also includes noise from zodiacal light, and thermal emission from the telescope, sunshade, and instrument.  Our direct imaging simulations use the noise simulator described in \citet{Robinson2016} for a 10-meter telescope. This coronagraph noise model simulates local and exo-zodiacal light, telescope thermal emission, dark current, read noise, and light leakage. It is highly customizable, allowing users to alter the distance to the planet-star system, the planet-star separation, the planet radius, the telescope diameter, the stellar spectrum, the telescope and instrument throughput, the inner and outer working angles, and the exposure time. Parameters for these values used here are the same as those described in \citet{Robinson2016} for a LUVOIR-class telescope. 

\subsection{Model Inputs}
The stellar spectra we use in our models include the Archean Sun (2.7 Ga), the modern Sun, AD Leo (M3.5V), GJ 876 (M4V), $\epsilon$ Eridani (K2V) and $\sigma$ Bo{\"o}tis (F2V). This stellar sample spans a range of activity levels, UV fluxes, and UV spectral slopes.

Our modern solar spectrum was modeled by \citet{Chance2010}, and our ``Archean'' Sun uses a modified spectrum based on the wavelength-dependent solar evolution correction from \citet{Claire2012} for 2.7 Ga.  This correction scales the absolute level of flux and accounts for the higher levels of solar activity, and therefore more UV radiation expected from a younger Sun. 

To test the impact of the UV spectrum of M dwarfs on haze generation, we compare results from two M dwarfs with different activity levels. M dwarfs can be highly active with frequent high-energy flares, although older M dwarfs may be more quiescent \citep{West2008}.  For a highly active flaring star, we use a time-averaged observed spectrum of AD Leo, an M dwarf with frequent flare events \citep{Hawley1991, Hunt-Walker2012}. Our AD Leo spectrum is discussed in \citet{Segura2005}. We also test haze generation using a spectrum for GJ 876, a known M4V planet host \citep{VonBraun2014}. The spectrum of GJ 876 is described in \citet{Domagal-Goldman2014} based on the spectrum reported in \citet{France2012}.

In addition to the M dwarfs, we use K2V ($\epsilon$ Eridani) and F2V ($\sigma$ Bo{\"o}tis) spectra described in \citet{Segura2003}. $\epsilon$ Eridani (3.2 pc) is a young star encircled by a dust ring \citep{Greaves1998} and is one of the closest known exoplanet hosts \citep{Hatzes2000}. $\epsilon$ Eridani is also chromospherically active \citep{Noyes1984}. $\sigma$ Bo{\"o}tis is an F2V star 15.5 pc away. 

\begin{figure*}[!htb]
\begin{center}
\includegraphics [scale=0.8]{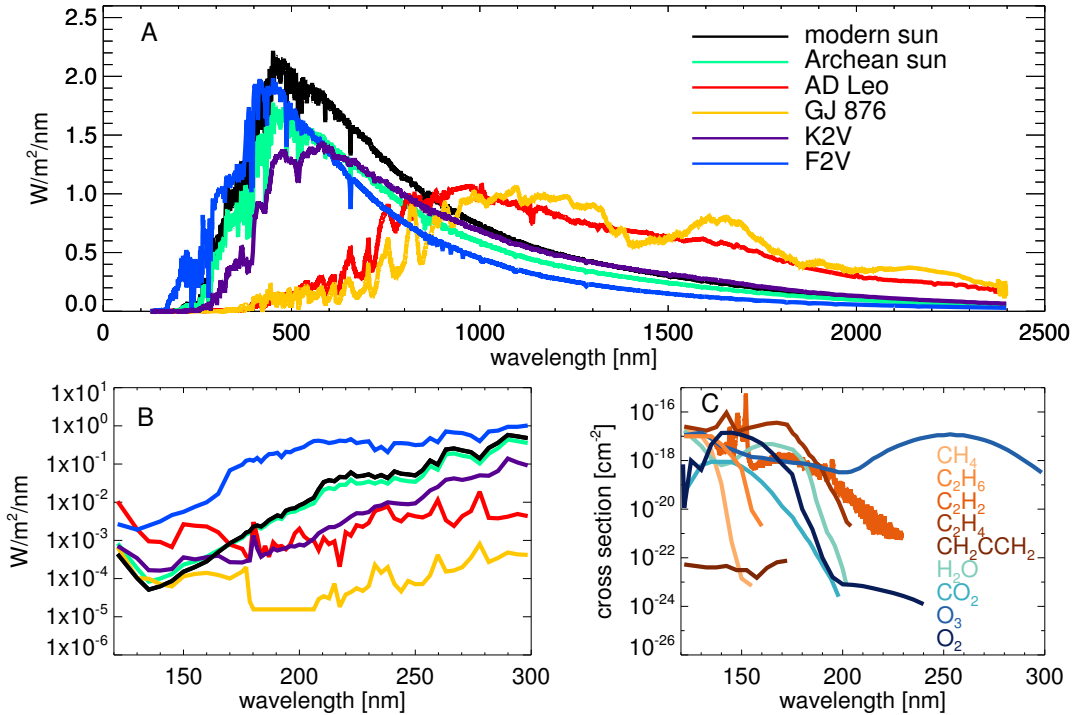}\\
\caption{Panel (A) displays all of the stellar spectra investigated by this study as received at the top of the planetary atmosphere.  Panel (B) zooms into the UV region of the stellar spectra.  In panel (C), UV cross sections of several interesting gases are shown with the same x-axis range as panel B.}
\label{fig:pod2_2}
\end{center}
\end{figure*}

For all stars except the modern Sun, we scale their total integrated fluxes to the solar constant for Earth at 2.7 Ga, which was 80\% less than the modern value (0.8 $\times$ 1360 W/m$^2$) to compare with our Archean results. Unlike the Archean Sun, the other stars do not include a wavelength-dependent stellar evolution correction. Figure \ref{fig:pod2_2} shows a comparison of the stellar spectra in the UV, visible, and near infrared (NIR) together with the UV cross-sections of several important gases. The plot of UV stellar spectra shows the actual resolution of the wavelength grid used by the photochemical model. These stellar spectra at full resolution are available for download on the VPL Spectral Database\footnote{https://depts.washington.edu/naivpl/content/spectral-databases-and-tools}. 

\begin{figure}[!htb]
\begin{center}
\includegraphics [scale=0.8]{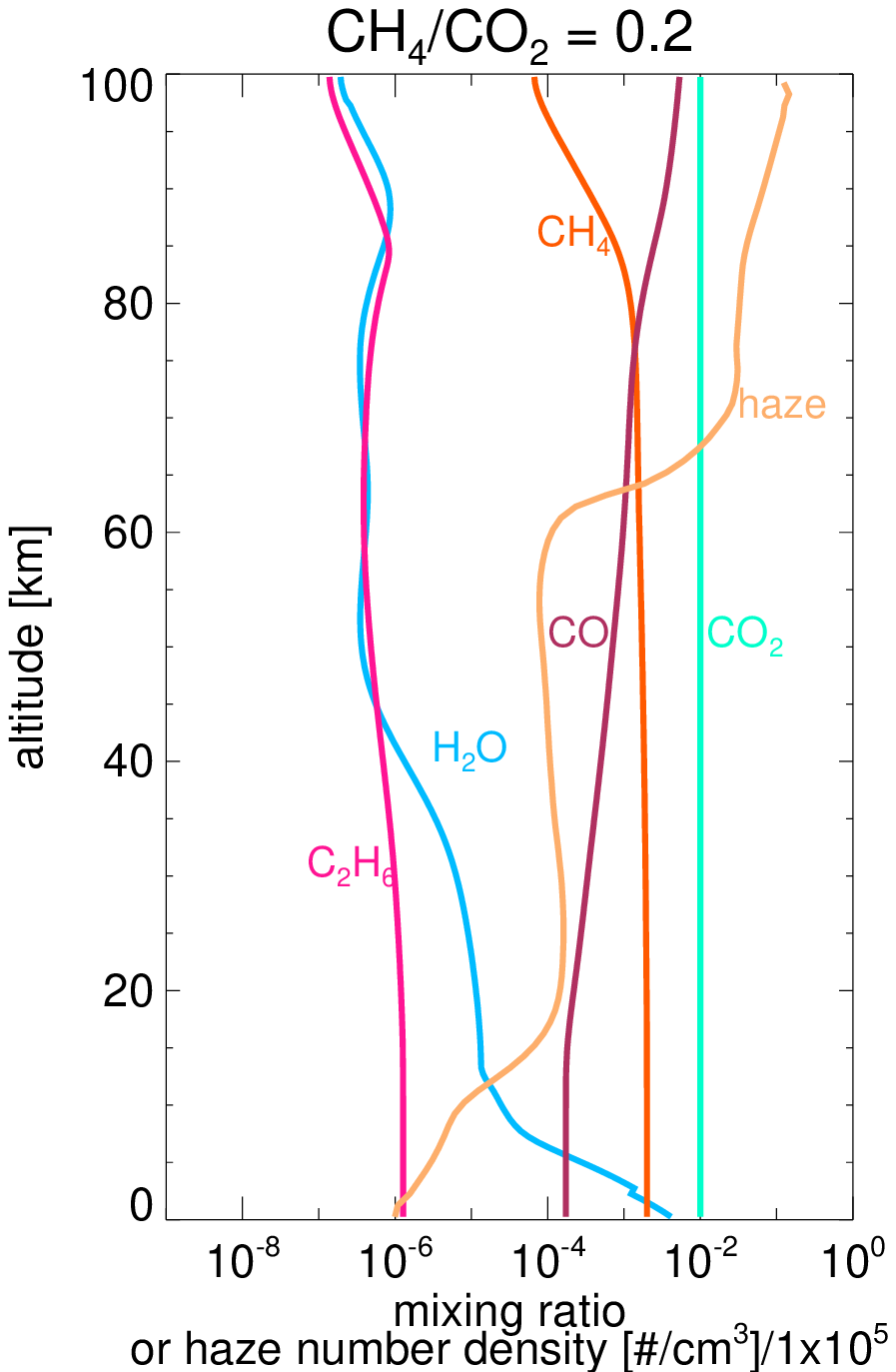}\\
\caption{The gas profiles for the nominal atmosphere investigated by this study for CH$_4$/CO$_2$ = 0.2 and pCO$_2$ = 0.01 bar for a planet orbiting the Archean sun (2.7 Ga).}
\label{fig:pod2_0}
\end{center}
\end{figure}

A total surface pressure of 1 bar is assumed in all situations. The nominal results presented here are for pCO$_2$ = 0.01 bar, and CH$_4$/CO$_2$ = 0.2 (Figure \ref{fig:pod2_0}). Note that the CH$_4$/CO$_2$ ratios we refer to apply to the planetary surface because CH$_4$ does not follow an isoprofile in the atmospheres we simulate. CO$_2$, on the other hand, is well-mixed. This CO$_2$ level is consistent with the paleosol measurements of \citet{Driese2011}, and this CH$_4$/CO$_2$ ratio is sufficient to form organic haze on Archean Earth. Molecular oxygen (O$_2$) is set at a mixing ratio of 1$\times$10$^{-8}$, corresponding to a time after the origin of oxygenic photosynthesis but prior to oxygen accumulation in the atmosphere. Haze particles are treated as fractals composed of 0.05 $\mu$m-sized spherical monomers, similar to the size of the monomers in Titan's hazes \citep{Rannou1997, Tomasko2008} and the same size of the monomers used by \citet{Wolf2010} in their study of fractal haze on Archean Earth. Haze scattering properties are derived using the optical constants of \citet{Khare1984} through the fractal mean field approximation \citep{Botet1997}. The \citet{Khare1984} optical constants were measured for Titan simulant hazes, but Archean-analog haze optical constants have only been measured at one wavelength (532 nm) in a previous study \citep{Hasenkopf2010}. The \citet{Khare1984} haze optical constants produce a reasonable match to the Hasenkopf haze measurement, and an extended discussion of our choice of optical constants can be found in our previous study \citep{Arney2016}.

We use the HITRAN 2012 linelists to generate our spectra \citep{Rothman2013}. The solar zenith angle is set at 60$^\circ$ for the direct imaging spectra, which approximates the average incoming solar flux at quadrature. As in $\mathtt{Atmos}$, we use fractal particles with scattering, absorption, and extinction efficiencies generated with the mean field approximation \citep{Botet1997} for our $\mathtt{SMART}$ simulations, and we use the optical constants from \citet{Khare1984}.

\section{Results}
In this section, we explore the factors affecting the formation of organic haze on Archean-analog planets orbiting stars of different spectral types, including the modern Sun. We then generate the spectra for the resultant planets. The strong UV and blue-wavelength absorption feature created by the haze may provide a UV shield for life on planetary surfaces, and we consider the strength of such a shield for each of the planets we simulate. Lastly, we consider the detectability of hazy spectral features using instrument simulators for JWST and a 10-m LUVOIR telescope. 

\subsection{Haze Formation and Surface Temperature Around Different Stellar Types}
To explore the effect of the stellar UV spectrum on haze formation, we ran $\mathtt{Atmos}$ to obtain chemically and climatically self-consistent results for Archean-like atmospheres under the influence of different host star spectra. Most of these models were run with CH$_4$/CO$_2$ = 0.2.  In the case of AD Leo and the K2V star, which did not form hazes with CH$_4$/CO$_2$ = 0.2, we ran additional atmospheres with higher CH$_4$/CO$_2$ ratios until a haze formed. These additional simulations have CH$_4$/CO$_2$ = 0.9 and 0.3, respectively. The F2V star did not form hazes at any CH$_4$/CO$_2$ ratios tested here (up to CH$_4$/CO$_2$ = 2). 

The compositions of our planets' atmospheres are strongly influenced by their host star's spectrum despite equivalent gaseous surface boundary conditions, underscoring the importance of photochemistry in exoplanet atmospheres. Table \ref{tab:pod2_1} shows the diurnally averaged stellar UV fluxes incident on the planet for near-UV (NUV, 300-400 nm), mid-UV (MUV, 200-300 nm), far-UV (FUV, 130-200 nm), and our photochemical model's Lyman alpha bin for each star. We also define and show ``interval 1'' (I1) as wavelengths between 120-140 nm and ``interval 2'' (I2) as wavelengths between 140-160 nm. I1 corresponds to the peak of the CH$_4$ UV cross section, and I2 to the peak of the CO$_2$ UV cross section. Note that a star's activity level (including Lyman alpha emission and extreme UV flux) is affected by a number of parameters including the stellar age and rotation rate. For instance, younger stars tend to have higher activity levels \citep[e.g.][]{West2008}. Stars that produce higher levels of FUV radiation compared to NUV and MUV tend to generate larger quantities of haze-destroying oxygen radicals because FUV can dissociate CO$_2$ and H$_2$O (Section \ref{sec:pod2_rxns}).  An exception is GJ 876, which also has a higher proportion of FUV relative to NUV and MUV; however GJ 876 has a lower absolute level of UV flux at these wavelengths. A test scaling GJ 876's total amount of UV radiation to the total level of UV radiation produced by AD Leo diminishes GJ 876's haze production rate. Similarly, a test decreasing AD Leo's total UV flux to that of GJ 876 increases its haze production rate. Consequently both the slope of the incident radiation (ratio of FUV to NUV or MUV) and the overall intensity of that radiation appear to affect haze production. 

The best predictor of whether a planet forms haze in our photochemical scheme appears to be the absolute level of flux in the I2 bin: here, the F2V star and AD Leo produce the most and second most flux, respectively, and these stars are least and second least efficient at haze formation as we discuss below. The K2V star has the third most amount of flux in the I2 bin, and while it is able to form a haze, it requires a higher CH$_4$/CO$_2$ ratio to do so compared to the stars with the lower I2 levels. GJ 876 has the lowest amount of flux in I2, and as we will show below, it is also the star hazes most easily form around. This suggests that oxygen species produced by CO$_2$ photolysis are the primary haze destroyers.

A well-characterized UV spectrum of the host star and good constraints on the planet's orbit will be important considerations for predicting incident UV radiation on a planet, the generation of hazes using photochemical models, and placing general constraints on photochemistry and climate of a planet.

Table \ref{tab:pod2_2} presents a comparison of the total integrated column densities of key gases in the atmospheres of our simulated environments, including some of the hydrocarbons involved in haze formation and oxygen radicals involved in destroying hydrocarbons. The values presented in this table are divided by the nominal total integrated column density for Archean Earth (with CH$_4$/CO$_2$  = 0.2) at 2.7 Ga, and the diversity of gas abundances for each star clearly illustrate how photochemistry impacts these atmospheres. In this table, C$_4$H$_2$ and C$_5$H$_4$ are direct precursors to haze particles according to our simplified haze formation scheme as discussed above \citep{Pavlov2001}. Ethane (C$_2$H$_6$), also shown, forms from photochemical reactions involving CH$_4$ and may be important for warming organic-rich atmospheres \citep{Haqq-Misra2008}.  

\begin{table*}[]
\centering
\caption{The diurnally averaged UV fluxes from these stars at the top of the atmosphere for NUV (300-400 nm), MUV (200-300 nm), FUV (130-200 nm) in W/m$^2$. Also included is the flux in our photochemical model's Lyman alpha (Ly$\alpha$) bin, the flux in ``interval 1'' (I1, 120-140 nm), and ``interval 2'' (I2, 140-160 nm) in W/m$^2$. The last three columns show FUV/MUV, FUV/NUV, and I1/I2 for each star.}
\label{tab:pod2_1}
\begin{tabular}{llllllllll}
Star        & NUV & MUV & FUV & Ly$\alpha$  & I1  & I2 & FUV/MUV & FUV/NUV & I1/I2\\
\hline
\hline
Modern Sun  & 44.2       & 7.01       & 0.047  & 0.0012 & 0.0015       & 0.0014    &    0.0067   & 0.0011 & 1.07 \\
Archean Sun & 31.2       & 4.79       & 0.041  & 0.0021   & 0.0027       & 0.0019  & 0.0088  & 0.0013 & 1.42 \\
AD Leo      & 0.42       & 0.17       & 0.063   & 0.033   & 0.033       & 0.017     & 0.37    & 0.15    & 1.94 \\
GJ 876      & 0.51       & 0.0095       & 0.0020 &0.0020  & 0.0021       & 0.00089     & 0.21       & 0.0039    &  2.35 \\
K2V         & 15.2       & 1.27       & 0.014  & 0.0028  &   0.0033     & 0.0024     & 0.011    & 0.0009  & 0.97 \\
F2V         & 57.7       & 23.1       & 2.1  & 0.012  &   0.022     & 0.052    & 0.092   & 0.03  & 0.42
\end{tabular}
\end{table*}

\begin{table*}[]
\centering
\caption{This table contains ratios of the total integrated column densities of gases in the atmospheres of Archean-analog planets around different spectral types divided by the total integrated column densities of gases for Archean Earth orbiting the Sun. A value of 1 indicates a column density identical to our nominal Archean Earth atmosphere. These planets have CH$_4$/CO$_2$= 0.2 except ``AD Leo - haze'' which has CH$_4$/CO$_2$= 0.9 and ``K2V - haze'' which has CH$_4$/CO$_2$= 0.3. Note C$_4$H$_2$ and C$_5$H$_4$ are direct precursors to hydrocarbon haze particles \citep{Pavlov2001}. }
\label{tab:pod2_2}
\begin{tabular}{lllllllll}
Star             & O     & O$_2$    & O$_3$    & OH     & NO &C$_2$H$_6$ & C$_4$H$_2$       & C$_5$H$_4$       \\
\hline
\hline
Modern Sun       & 0.34  & 0.76  & 0.26  & 1.29   & 1.74 & 2.46 & 1.39       & 2.11       \\
AD Leo - no haze & 27.05 & 6.95  & 0.1   & 14.82 & 3.15 & 5    & 1.86x10$^{-5}$  & 1.00x10$^{-6}$  \\
AD Leo - haze    & 12.69 & 2.15  & 0.022 & 3.24  & 4.21 & 19.2 & 0.069      & 0.024      \\
GJ 876           & 0.076 & 0.68  & 0.33  & 1.1 & 10.6  & 1.74 & 1.25       & 5.21       \\
K2V - no haze    & 1.31  & 2.85  & 0.71  & 2.6 & 0.71   & 3.37 & 3.84x10$^{-4}$  & 4.41x10$^{-4}$  \\
K2V - haze       & 0.5   & 2.66  & 0.31  & 0.99 & 1.65  & 2.33 & 1.28       & 5.94       \\
F2V              & 78.64 & 265.5 & 169.4 & 34.11 & 0.19 & 1.74 & 9.11x10$^{-11}$ & 1.02x10$^{-12}$
\end{tabular}
\end{table*}

In addition to the gas profiles, $\mathtt{Atmos}$ was also used to calculate the temperature profiles for the simulations shown in Table \ref{tab:pod2_2}. The results of these climate calculations are provided in Table \ref{tab:pod2_3}. The diversity of surface temperatures in this table is due to the climatic effects of different haze thicknesses, different greenhouse gas abundances, and the host star spectral energy distribution. These effects are discussed in detail in the sections below. 

\begin{table*}[]
\centering
\caption{The surface temperatures of planets orbiting each spectral type for CH$_4$/CO$_2$= 0.2 except ``AD Leo - haze'' which has CH$_4$/CO$_2$= 0.9 and ``K2V - haze'' which has CH$_4$/CO$_2$= 0.3. We also show the top-of-atmosphere planetary albedo, incoming shortwave radiation, outgoing shortwave radiation, and outgoing longwave radiation.}
\label{tab:pod2_3}
\begin{tabular}{llllll}
Star           & Surface  & Planetary  & Incoming  & Outgoing  & Outgoing  \\
           &  Temp &  Albedo & shortwave  & shortwave & longwave \\
               &                     &                  &      (W/$m^2$)       & (W/$m^2$)       &  (W/$m^2$)     \\
               \hline
\hline
Modern Sun     & 299 K               & 0.216            & 342                                 & 74.6                         & 267                                \\
Archean Sun    & 272 K               & 0.238            & 278                                 & 66.4                         & 212                                \\
AD Leo - no haze         & 310 K               & 0.087            & 278                                 & 24.2                         & 253                                \\
AD Leo - haze & 317 K               & 0.067            & 278                                 & 18.7                         & 259                                \\
GJ 876         & 301 K               & 0.137            & 278                                 & 38.3                         & 240                                \\
K2V - no haze           & 297 K               & 0.202            & 278                                 & 56.4                         & 221                                \\
K2V - haze    & 282 K               & 0.210             & 278                                 & 58.7                         & 219                                \\
F2V            & 277 K               & 0.322            & 278                                 & 89.6                         & 188                               
\end{tabular}
\end{table*}

\begin{figure*}[!htb]
\begin{center}
\includegraphics [scale=0.8]{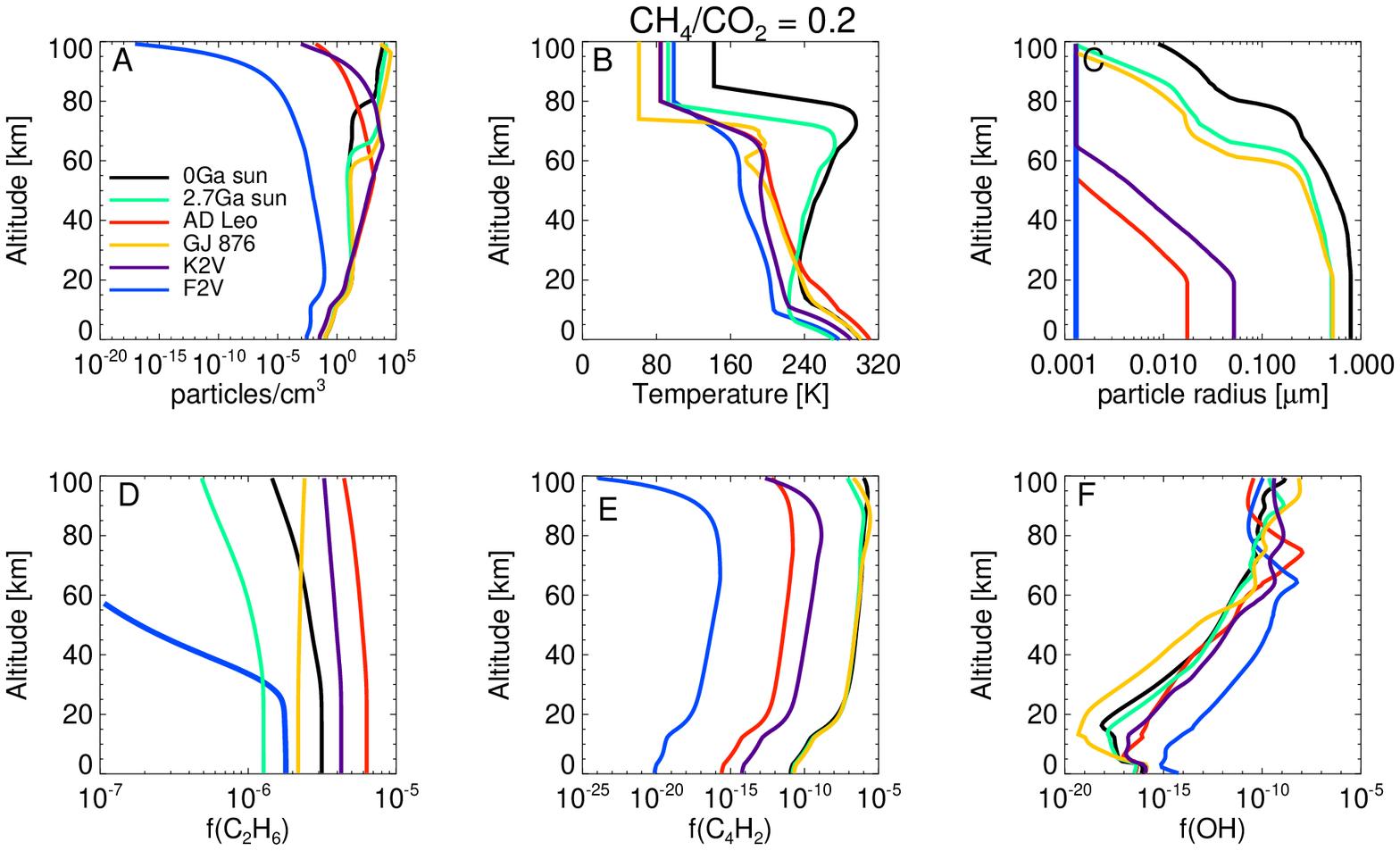}\\
\caption{Displayed here are: A) the number density of haze particles, B) the temperatures of the CH$_4$/CO$_2$ = 0.2 atmospheres, C) radii of haze particles, D) the C$_2$H$_6$ mixing ratios, E) the mixing ratio of C$_4$H$_2$ particles, which is the primary vector that condenses directly out to form aerosols in our chemical scheme, and F) the mixing ratio of OH radicals, to illustrate the buildup of such oxygen species in the atmosphere of the F star.}
\label{fig:pod2_3}
\end{center}
\end{figure*}

To illustrate these atmospheres' gas, haze, and temperature profiles, these quantities are shown for the CH$_4$/CO$_2$ = 0.2 planets in Figure \ref{fig:pod2_3}. The profiles for the nominal Archean environment are shown with the green lines for comparison. 

\subsection{Causes of different rates of haze formation around different stars}\label{sec:pod2_rxns}
The photochemical production of oxygen-bearing gases has an important impact on the ability of each atmosphere simulated here to form haze. The generation of O from CO$_2$ photolysis (and photolysis of other O-bearing species such as NO$_2$ and H$_2$O) leads to the creation of oxidized species (including O itself) that can react with  organics on the reaction pathway to haze formation. 

Table \ref{tab:pod2_6} shows a comparison of the principle photolysis reactions involving CO$_2$, H$_2$O, and NO$_2$ in each atmosphere that produce hydrocarbon-consuming oxygen species. An asterisk marks the fastest reaction in each atmosphere. CO$_2$ photolysis is the most efficient source of O and O$^1$D in every atmosphere. The reaction rates scale with the amount of UV flux able to dissociate a given species. For instance, compare the reaction rates in Table \ref{tab:pod2_6} with the UV fluxes and cross sections in Figure \ref{fig:pod2_2}: stars with elevated fluxes at the wavelengths overlapping the UV cross sections of these O-producing species produce higher amounts of oxygen species through photolysis. In general, the more oxygen an atmosphere produces from photolysis of species like H$_2$O, CO$_2$, and NO$_2$, the thinner the hazes. Thus, to predict whether a star is likely to have a planet with organic haze in the habitable zone, we will require measurements of the shape of its UV spectrum, especially for wavelengths controlling CO$_2$ photolysis, which is the dominant predictor of haze destruction.

\begin{table}[]
\centering
\caption{Column integrated rates (reactions/sec) for photolysis of H$_2$O, CO$_2$, and NO$_2$ in the atmospheres of the CH$_4$/CO$_2$ = 0.2 planets around each star. Also shown is the column integrated rate of all CH$_4$ photolysis reactions.}
\label{tab:pod2_6}
\begin{tabular}{lll}\
Star        & Oxygen-producing  & Rates      \\
           & photolysis reactions &       \\
           \hline
           \hline
Modern Sun  & *CO$_2$ + h$\nu$ $\rightarrow$CO + O        & 3.12x10$^{11}$  \\
            & CO$_2$ + h$\nu$ $\rightarrow$CO + O$^{1}$D       & 2.02x10$^{11}$  \\
            & H$_2$O + h$\nu$ $\rightarrow$H + OH         & 7.024x10$^{10}$ \\
            & NO$_2$ + h$\nu$ $\rightarrow$NO+ O          & 1.52x10$^{11}$  \\
            & all CH$_4$ photolysis					    & 1.31x10$^{11}$ \\
Archean Sun & *CO$_2$ + h$\nu$ $\rightarrow$CO + O        & 3.87x10$^{11}$  \\
            & CO$_2$ + h$\nu$ $\rightarrow$CO + O$^{1}$D       & 3.38x10$^{11}$  \\
            & H$_2$O + h$\nu$ $\rightarrow$H + OH         & 5.65x10$^{10}$  \\
            & NO$_2$ + h$\nu$ $\rightarrow$NO+ O          & 1.40x10$^{11}$  \\
            & all CH$_4$ photolysis					    & 2.14x10$^{11}$ \\            
AD Leo      & *CO$_2$ + h$\nu$ $\rightarrow$CO + O        & 2.19x10$^{12}$  \\
            & CO$_2$ + h$\nu$ $\rightarrow$CO + O$^{1}$D       & 2.51x10$^{11}$  \\
            & H$_2$O + h$\nu$ $\rightarrow$H + OH         & 5.78x10$^{11}$  \\
            & NO$_2$ + h$\nu$ $\rightarrow$NO+ O          & 6.84x10$^{10}$  \\
            & all CH$_4$ photolysis					    & 5.97x10$^{11}$ \\
GJ 876      & CO$_2$ + h$\nu$ $\rightarrow$CO + O        & 4.17x10$^{9}$   \\
            & *CO$_2$ + h$\nu$ $\rightarrow$CO + O$^{1}$D       &1.18x10$^{11}$   \\
            & H$_2$O + h$\nu$ $\rightarrow$H + OH         & 3.76x10$^{10}$  \\
            & NO$_2$ + h$\nu$ $\rightarrow$NO+ O          & 4.37x10$^{10}$  \\
            & all CH$_4$ photolysis					    & 9.57x10$^{10}$ \\
K2V         & *CO$_2$ + h$\nu$ $\rightarrow$CO + O        & 3.69x10$^{11}$  \\
            & CO$_2$ + h$\nu$ $\rightarrow$CO + O$^{1}$D       & 3.42x10$^{11}$  \\
            & H$_2$O + h$\nu$ $\rightarrow$H + OH         & 9.21x10$^{10}$  \\
            & NO$_2$ + h$\nu$ $\rightarrow$NO+ O          & 1.2x10$^{11}$   \\
             & all CH$_4$ photolysis					    & 1.20x10$^{11}$ \\
F2V         & *CO$_2$ + h$\nu$ $\rightarrow$CO + O        & 1.05x10$^{14}$ \\
            & CO$_2$ + h$\nu$ $\rightarrow$CO + O$^{1}$D       & 7.774x10$^{12}$ \\
            & H$_2$O + h$\nu$ $\rightarrow$H + OH         & 8.22x10$^{12}$  \\
            & NO$_2$ + h$\nu$ $\rightarrow$NO+ O          & 1.31x10$^{11}$  \\        
             & all CH$_4$ photolysis					    & 2.43x10$^{11}$ \\

\end{tabular}
\end{table}

\begin{figure*}[!htb]
\begin{center}
\includegraphics [scale=0.65]{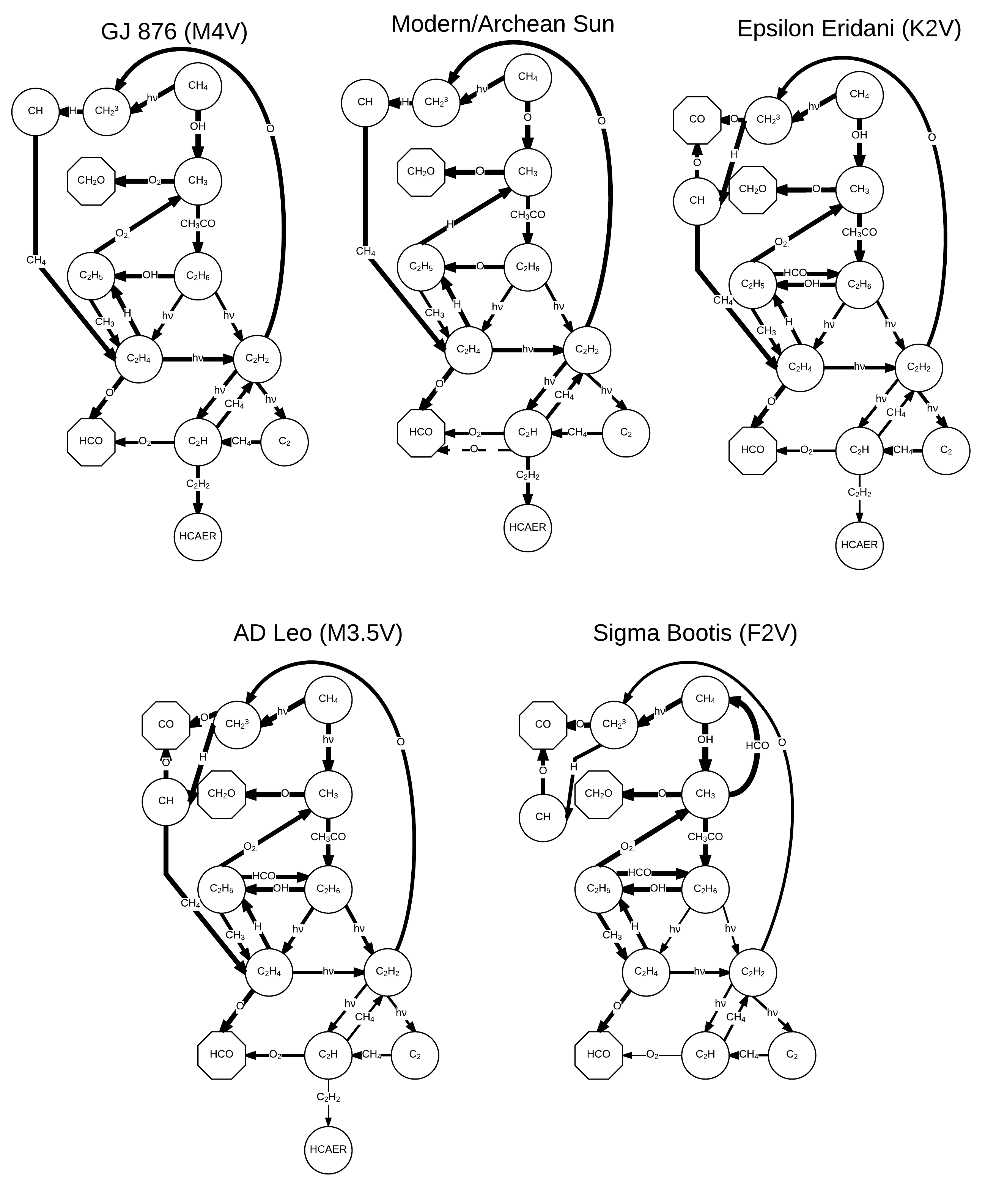}\\
\caption{The fastest reaction network towards haze formation (HCAER) and the fastest hydrocarbon sinks for each planet. The Archean and modern sun reaction networks are the same except for the reaction forming HCO from C$_2$H: the dominant reaction involves O$_2$ for the modern sun (solid arrow) and O for the Archean sun (dashed arrow). The F2V star, on the other hand, has a complex network in which hydrocarbons efficiently react with oxygen species, and haze is not formed. Thicker arrows indicate faster reaction rates.}
\label{fig:pod2_7pt5}
\end{center}
\end{figure*}

%%figure
Figure \ref{fig:pod2_7pt5} shows the hydrocarbon chemical reaction network with the fastest reactant rates that lead to haze production or haze sinks for each star. ``HCAER'' in this figure stands for the hydrocarbon species that most efficiently condenses out as haze particles, C$_4$H$_2$. Species outlined by hexagons represent major sinks of haze-forming gases, and essentially a truncation of the haze-forming reaction network.  The dominant overall pathway to haze formation for the haze-forming planets examined here is:

\begin{chequation}
\begin{align}
\textrm{CH}_4 \textrm{+ h}\nu \rightarrow \textrm{CH}_2^3 + \textrm{H + H}  \\  \label{ch_net} 
\textrm{CH}_2^3 + \textrm{H} \rightarrow \textrm{CH + H}_2 \\
\textrm{CH + CH}_4 \rightarrow \textrm{C}_2\textrm{H}_4 \textrm{+ H} \\
\textrm{C}_2\textrm{H}_4 \textrm{+ h}\nu \rightarrow \textrm{C}_2\textrm{H}_2 \textrm{+ H + H} \\
\textrm{C}_2\textrm{H}_2 +\textrm{h}\nu \rightarrow \textrm{C}_2\textrm{H + H} \\
\textrm{C}_2\textrm{H + C}_2\textrm{H}_2 \rightarrow \textrm{C}_4\textrm{H}_2\textrm{ + H}
\end{align}
\end{chequation}

In this pathway, CH needs to react with CH$_4$ to form C$_2$H$_4$, but in atmospheres that generate large quantities of oxygen species, CH can instead readily react with O to form CO via CH + O $\rightarrow$ CO + H. Once C$_2$H$_4$ forms, it can exit the haze-formation path if it reacts with O to form HCO, or it can continue on the haze-formation path if it is photolyzed to form C$_2$H$_2$. Then, C$_2$H$_2$ can be photolyzed to produce C$_2$H, but if C$_2$H$_2$ instead reacts with O in an oxygen-rich atmosphere, it can form CH$_2^3$; although CH$_2^3$ is involved in haze formation, going from C$_2$H$_2$ to  CH$_2^3$ does not advance further in the haze formation scheme towards higher order hydrocarbons. If the C$_2$H produced from C$_2$H$_2$ reacts with O or O$_2$, it will result in HCO, which is not useful for haze formation.

Once the reaction network has produced the gases needed to form C$_4$H$_2$ (HCAER), it condenses out as haze particles, via C$_2$H$_2$ + C$_2$H $\rightarrow$ C$_4$H$_2$ + H. Alternatively, C$_2$H$_2$ can react with CH to form C$_3$H$_2$, which begins a second chain of polymerizations that can lead to C$_5$H$_4$, which also condenses out as haze particles (HCAER2). However, this is less efficient than the process forming C$_4$H$_2$ and is not shown in Figure \ref{fig:pod2_7pt5} or in the reaction network outlined above.

Table \ref{tab:tableofhell} shows the ratios of the total integrated reaction rates that remove hydrocarbons from the haze formation chain (via reaction with oxygen radicals) versus the reactions that step toward haze particle formation. The reactions in the table represent key stages of the dominant haze-formation process outlined above. There is a clear difference between the planets that form haze and those that do not. The haze-poor planets generally favor reactions with oxygen radicals over reactions that lead to haze formation. As discussed previously, the primary driver of this difference in behavior is the amount of flux in the 140-160 nm region, where CO$_2$ is most efficiently photolyzed because oxygen sourced from CO$_2$ photolysis is the primary source of haze-precursor destruction, although the other oxygen-bearing gases contribute as well.

An alternative, but less efficient, way of forming C$_2$H$_4$ involves the formation of ethane. This haze-formation pathway requires production of the methyl radical (CH$_3$). Although oxygen radicals can frustrate haze formation later on in the reaction network, they are initially helpful in forming CH$_3$. For every star but AD Leo, the most efficient vectors towards forming CH$_3$ are: 

\begin{chequation}
\begin{align}
\textrm{CH}_4 \textrm{+ O} \rightarrow \textrm{CH}_3 \textrm{+ OH}   \label{ch_1}\\ 
\textrm{CH}_4 \textrm{+ OH} \rightarrow \textrm{CH}_3 \textrm{+ H}_2\textrm{O}\label{ch_2} 
\end{align}
\end{chequation}

AD Leo, meanwhile, most efficiently forms CH$_3$ from CH$_4$ photolysis due to its high Lyman-$\alpha$ output overlapping with the peak of the CH$_4$ UV cross section, but this is only a factor of 1.15 times faster than CH$_4$ + OH. 

Once CH$_3$ forms, an exit from the haze-formation network occurs if it reacts with O to form formaldehyde, CH$_2$O, which ultimately ends up as CO$_2$:

\begin{chequation}
\begin{align}
\textrm{CH}_3 \textrm{+ O} \rightarrow \textrm{CH}_2\textrm{O + H} \label{ch_3}
\end{align}
\end{chequation}

The above oxidation of CH$_3$ competes with the formation of ethane (C$_2$H$_6$) via reaction of CH$_3$ with another CH$_3$, or more commonly with CH$_3$CO (produced from CH$_3$ + CO):

\begin{chequation}
\begin{align}
\textrm{CH}_3 \textrm{+ CH}_3 \rightarrow \textrm{C}_2\textrm{H}_6 \\
\textrm{CH}_3\textrm{CO + CH}_3 \rightarrow \textrm{C}_2\textrm{H}_6\textrm{ + CO}
\end{align}
\end{chequation}

A number of reactions can then occur with C$_2$H$_6$ that are relevant to haze formation. Ethane can react with oxygen radicals to form C$_2$H$_5$. Then, C$_2$H$_5$ can either react with H or O$_2$ to re-form CH$_3$, or it can react with CH$_3$ to advance towards C$_2$H$_4$. Alternatively, C$_2$H$_6$ can be photolyzed to form C$_2$H$_4$ directly (and, less efficiently, C$_2$H$_2$). Once C$_2$H$_4$ forms, haze formation advances towards C$_4$H$_2$ (HCAER) through the same steps outlined above in the dominant haze formation network after C$_4$H$_2$ is formed.

\begin{table}[]
\centering
\caption{In this table, ``Ratio X'' (where X is either CH, C$_2$H$_4$, C$_2$H$_2$ or C$_2$H) represents the haze network truncation ratio, which is the ratio of the total integrated reaction rates of X with oxygen species (thus frustrating the haze-formation process) to the total integrated reaction rates of X that step towards haze formation. A horizontal line separates the stars that form haze at CH$_4$/CO$_2$ = 0.2 (above the line) from the ones that do not (below the line). Ratios exceeding unity mean that reactions with oxygen species are more efficient than reactions towards haze particle formation. Reactions with oxygen species are not always less efficient than reactions leading towards haze formation even in the hazy atmospheres (e.g., for C$_2$H$_4$ and C$_2$H$_2$), but they are markedly faster in the haze-free atmospheres compared to the hazy ones.}
\label{tab:tableofhell}
\begin{tabular}{lllll}
Star        & Ratio & Ratio  & Ratio  & Ratio \\
          & CH & C$_2$H$_4$  & C$_2$H$_2$  & C$_2$H \\
        \hline
       \hline
GJ 876                 & 0.12 & 1.88                              & 1.7                                      & 0.07                                        \\
Modern sun          & 0.15 & 3.22                              & 2.48                                     & 0.13                                        \\
Archean                 & 0.29 & 4.92                              & 2.8                                      & 0.45                                        \\
\hline
K star               & 1.45 & 86                                & 10                                       & 60                                          \\
AD Leo             & 1.78 & 121                               & 42                                       & 703                                         \\
F star                & 16.2 & 60                                & 7.5                                      & 1.51$\times10^{6}$                                                                         
\end{tabular}
\end{table}

Below, we present an analysis of haze formation and its climatic consequence for each host star type compared to our nominal Archean results to explore the differing atmospheric compositions and temperatures of these worlds. 

\subsection{Hazes with the Archean solar constant}
We present a detailed discussion of haze formation for Archean Earth orbiting the sun at 2.7 billion years ago in \citet{Arney2016}. Haze formation for pCO$_2$ = 0.01 begins to noticeably impact the Earth's spectrum at CH$_4$/CO$_2$ = 0.18. The temperature of the planet's surface drops from $\sim$284 K when no haze is present to 272 K when a haze is in place at CH$_4$/CO$_2$ = 0.2. Although this surface temperature is below the freezing point of water, 3D climate studies have suggested planets like early Earth with global average temperatures down to 250 K can still maintain stable open ocean waters near the equator \citep{Charnay2013}, so this cold temperature can still be considered ``habitable'' as the planet could still support liquid water at the surface.  

\subsubsection{Hazes with the modern solar constant}
For the planet orbiting the modern (0 Ga) Sun, we find that larger haze particles form in the atmosphere of this hotter planet when compared to the planet experiencing the Archean solar constant (80\% of modern). Particle coagulation proceeds more efficiently in hotter atmospheres, leading to larger particles \citep{Arney2016}. For CH$_4$/CO$_2$ = 0.2, the 0 Ga planet has a surface temperature of 299 K and a maximum haze particle radius of 0.79 $\mu$m. By comparison, the 2.7 Ga, planet has a surface temperature of 272 K and a maximum haze particle radius of 0.51 $\mu$m

Haze formation is also more efficient in the modern planet's atmosphere because the present day solar spectrum is less active at UV wavelengths shorter than $\sim$150 nm (Figure \ref{fig:pod2_3}) when compared to the early Sun. Therefore, it tends to generate smaller quantities of the types of oxygen species that destroy hydrocarbon haze precursors from H$_2$O and CO$_2$ photolysis (Table \ref{tab:pod2_2}, and see Section \ref{sec:pod2_rxns} for a discussion of these processes). 

\subsubsection{M dwarfs}\label{sec:pod2_mdwarfs}

AD Leo outputs considerable UV flux, but we find that it is also inefficient at generating hazes compared to the Archean and modern Sun. A scant haze of small particles (maximum particle radius = 0.017 $\mu$m) is present at CH$_4$/CO$_2$ = 0.2 and is spectrally indistinguishable from a world without haze (Section \ref{sec:pod2_spectra}). The CH$_4$/CO$_2$ ratio must reach 0.9 before AD Leo's haze begins to alter the spectrum, but even then, the impact is small as discussed in Section \ref{sec:pod2_spectra}. 

AD Leo's inability to efficiently generate hydrocarbon haze is a result of the relatively large quantities of oxygen radicals generated in the atmosphere of its planet from its high FUV flux (Table \ref{tab:pod2_2}). AD Leo is a highly active M dwarf and produces excess flux at $\lambda < 170$ nm compared to every other star considered except the F dwarf star. This spectral region is coincident to the peaks of the CO$_2$ and H$_2$O cross sections. Therefore, it is relatively efficient at photolyzing these gases to produce oxygen species that can destroy the higher order hydrocarbons necessary for haze formation (Section \ref{sec:pod2_rxns}). For an Archean analog orbiting AD Leo, the ``source'' for the higher-order hydrocarbons goes up for these higher UV fluxes, but not as quickly as the ``sink'' for these species. Because of this, we find that even at the CH$_4$/CO$_2$ ratios exceeding unity that were tested, the haze around the AD Leo planet remains optically thin in the UV. 

Our results for AD Leo seem to indicate that Earthlike planets around M dwarfs are unlikely to have organic haze, but this is not the case for GJ 876.  This star produces smaller amounts of hydrocarbon-destroying oxygen species due to its lower levels of UV radiation relative to every other star. Haze particles for the GJ 876 planet reach radii of 0.52 $\mu$m at CH$_4$/CO$_2$ = 0.2, similar to the size of the particles around the Archean Sun. Haze begins to noticeably alter the spectrum for the GJ 876 planet at CH$_4$/CO$_2$ = 0.12, which is a lower ratio than for the equivalent planet orbiting the Archean Sun, for which the spectral impact of haze begins to become apparent at about CH$_4$/CO$_2$= 0.18. In fact, GJ 876's planet exhibits the lowest CH$_4$/CO$_2$ratio able to form a haze among the stars tested here. Hazy atmospheres around planets orbiting M dwarfs like GJ 876 may therefore occur at lower CH$_4$/CO$_2$ratios than for other types of stars. 

The Archean-analog planets around the M dwarfs are warmer than the one orbiting around the Archean Sun despite having equivalent levels of total incident radiation at the top of their atmospheres.  For CH$_4$/CO$_2$= 0.2, T$_{surf}$ = 310 K for the AD Leo planet, and T$_{surf}$ = 301 K for the GJ 876 planet. These relatively high temperatures are caused by the three factors described below.

The first reason for the warm M dwarf planets, discussed in \citet{Kopparapu2013}, is that M dwarfs produce the bulk of their radiation in the near-infrared where Rayleigh scattering is weak and gaseous absorbers, particularly water vapor, have broad absorption features. These factors act to reduce the planetary albedo relative to a planet with the same atmospheric composition, but orbiting a Sun-like star (Table \ref{tab:pod2_3}). So, a planet around an M dwarf at the equivalent flux distance of a solar-type star will naturally produce warmer temperatures with an equivalent atmosphere. 

The second reason for the warm M dwarf temperatures is that these atmospheres contain large amounts of greenhouse gases. Notice from Table \ref{tab:pod2_2} that at CH$_4$/CO$_2$= 0.2, AD Leo is able to build up about 5$\times$ as much C$_2$H$_6$ than the nominal Archean Earth, and C$_2$H$_6$ is a greenhouse gas. Accordingly, this planet is warmer than the GJ 876 planet, which only has 1.74 $\times$ as much C$_2$H$_6$ as the nominal Archean planet. This finding is consistent with the results of \citet{Domagal-Goldman2011}, which determined that AD Leo builds up larger amounts of C$_2$H$_6$ than a planet orbiting the Sun. 

The final reason for the warm temperatures of M dwarf planets with haze is due to the spectral properties of the haze itself. The bulk of the M dwarf radiation arrives between 700 and 2500 nm (Figure \ref{fig:pod2_2}), but  fractal particle extinction efficiency decreases by 1-2 orders of magnitude in the NIR compared to the visible, so the haze is relatively transparent at these wavelengths (Figure \ref{fig:pod2_1}). Therefore, cooling from organic haze is far less relevant to planets orbiting M dwarfs than it is for other stellar types with bluer spectra. This is why the hazy AD Leo planet with CH$_4$/CO$_2$ = 0.9 is actually hotter (T$_{surf}$ = 317 K) than the haze-free AD Leo planet (T$_{surf}$ = 310 K); the hazy AD Leo planet has more methane and its haze does not effectively scatter the incident radiation back to space. Implications of these hazes' low NIR opacities are discussed in Section \ref{sec:pod2_cooling}.

%%%number chemical reactions!!%%
Although the hazy GJ 876 planet has a warmer surface temperature than the 0 Ga planet, the temperature feedbacks on particle size discussed in the context of the solar-type stars do not apply here because this star emits comparatively little UV for the haze to absorb and warm the upper atmosphere where particle coagulation proceeds. As can be seen in Figure \ref{fig:pod2_3}, GJ 876 does not show a prominent stratospheric temperature inversion that the 0 Ga and 2.7 Ga solar-type planets have.

\subsubsection{K2V Dwarf}
The K2V star has excess UV flux at wavelengths $<$ 170 nm compared to the Archean Sun, as does AD Leo (Figure \ref{fig:pod2_2}) -- although with about one order of magnitude lower flux. The K2V star has a relatively high level of both FUV flux and FUV/MUV, and it is able to produce oxygen radicals and unable to form a thick haze at CH$_4$/CO$_2$= 0.2. For CH$_4$/CO$_2$= 0.2, the K2V planet generates a sparse haze of very small particles (radius $<$ 0.05 $\mu$m) that produce a negligible spectral effect, while the Archean Earth haze particles are an order of magnitude larger.  However, it is efficient at forming haze at slightly higher CH$_4$/CO$_2$ratios. At CH$_4$/CO$_2$= 0.3, the particles reach a radius of 0.51 $\mu$m, similar to the size of the particles for the planets orbiting the Archean Sun and GJ 876 planet. 

For a planet with CH$_4$/CO$_2$= 0.2, the K2V planet has an average surface temperature of 297 K, and at CH$_4$/CO$_2$= 0.3, the average surface temperature drops by 15 K to 282 K due to the accumulation of haze and haze-induced cooling. This cooling is not as strong as for the hazy Archean planet orbiting the sun because the K dwarf spectrum is shifted slightly redward of the G dwarf spectrum where the haze is more transparent. Also, this hazy planet has more CH$_4$ than the corresponding hazy CH$_4$/CO$_2$ = 0.2 planet for the Archean sun so it is warmer. 

\subsubsection{F2V Dwarf}

We find that the F dwarf planet does not form a haze due to high incident UV flux. The F dwarf spectrum produces more UV flux than all the other stars we consider at all wavelengths except at Lyman $\alpha$ ($\lambda$ = 121.6 nm). Its high UV flux efficiently photodissociates hydrocarbon species and generates extremely large quantities of oxygen radicals (Table \ref{tab:pod2_2}) compared to the other stars (Section \ref{sec:pod2_rxns}). This is consistent with the previously noted ability of F stars to generate amounts of oxygen large enough to significantly impact photochemistry \citep{Domagal-Goldman2014}. It is not possible to generate hazes in the F2V planet's atmosphere even with CH$_4$/CO$_2$$>$1. This is because hydrocarbons are too efficiently destroyed in this type of atmosphere, although significantly more reducing conditions without CO$_2$ (i.e. more Titan-like) that would produce fewer oxygen radicals were not tested here.

The climate of the haze-free F star planet is relatively cool compared to other spectral types: the CH$_4$/CO$_2$= 0.2 planet has a mean surface temperature of 277 K despite its clear sky.  This low temperature is due to the F star spectral energy distribution peaking near 400 nm, a wavelength at which Rayleigh scattering from the planet's atmosphere efficiently reflects much of the incident energy back to space, so that a larger fraction of the F star incident radiation avoids NIR absorption bands \citep{Kopparapu2013}.  Therefore, at an equivalent flux distance, an F star planet would naturally be cooler than a planet orbiting a star with a redder spectrum.

\subsection{Spectra}\label{sec:pod2_spectra}
\begin{figure*}[!htb]
\begin{center}
\includegraphics [scale=0.7]{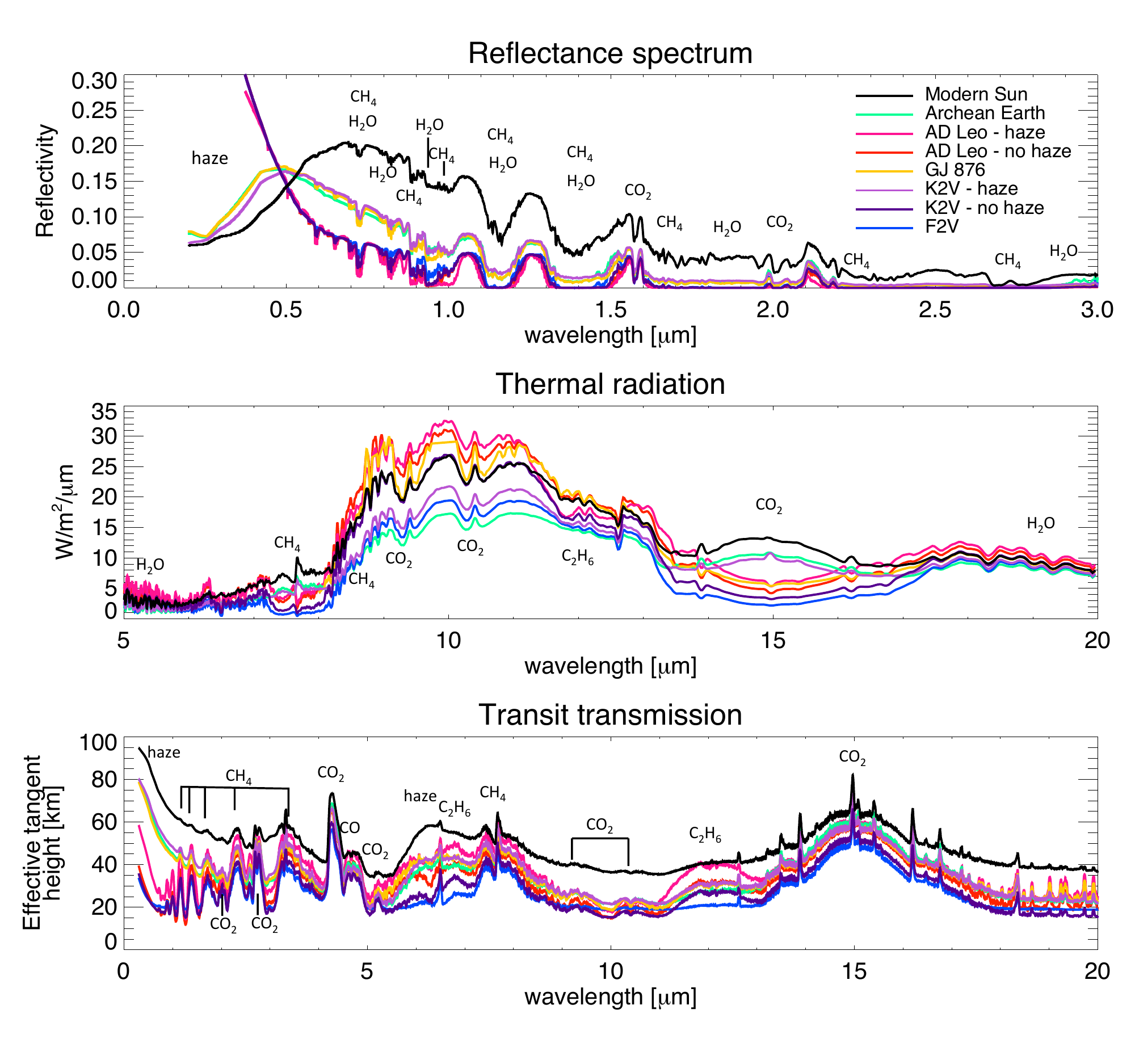}\\
\caption{Reflectance spectra (top panel), thermal radiation (middle panel) and transit transmission spectra (bottom panel) for the Archean Earth-type planets around varied spectral types.  The transit transmission spectra show the effective tangent height, which is the minimum altitude the atmosphere is transparent to as a function of wavelength for light traveling tangent to the planet surface. All spectra shown are for CH$_4$/CO$_2$= 0.2 except spectrum labeled ``K2V - haze'', which has CH$_4$/CO$_2$= 0.3, and the ``AD Leo - haze'' spectrum, which has CH$_4$/CO$_2$= 0.9.}
\label{fig:pod2_4}
\end{center}
\end{figure*}

Reflectance, thermal emission, and transit transmission spectra for the Archean-analog planets are presented in Figure \ref{fig:pod2_4} for all stellar spectral types studied here.  All of these planets have CH$_4$/CO$_2$= 0.2 except the spectrum labeled ``K2V - haze'', which has CH$_4$/CO$_2$= 0.3 and ``AD Leo - haze'' which has CH$_4$/CO$_2$= 0.9, the ratios required to form haze for these planets. At CH$_4$/CO$_2$= 0.2, planets around AD Leo, the K2V star, and the F2V star do not have spectrally apparent hazes in their atmospheres, but the Archean Sun, modern Sun, and GJ 876 planets do. As noted before, the F2V star does not generate organic hazes even at CH$_4$/CO$_2$ratios greater than unity. 

Haze absorbs strongly at blue and UV wavelengths, causing the reflectance spectra (top panel of Figure \ref{fig:pod2_4}) of the hazy worlds to have lower albedos at these wavelengths.  When the haze is thick enough to affect the spectrum, it creates a large absorption feature at these short wavelengths. Thus, rather than the Rayleigh scattering-induced increase in reflectivity at short wavelengths seen for the haze-free planets, hazy worlds produce their peak spectral brightness at visible wavelengths. The UV-blue haze absorption feature can be seen for the Archean and modern Sun, GJ 876, and hazy K2V planets, although the sparse haze around the CH$_4$/CO$_2$= 0.9 AD Leo planet is thin enough to be almost spectrally indistinguishable from a clear sky world (Figure \ref{fig:pod2_4}).

There are large spectral differences for the planet orbiting the modern sun compared to the hazy planets around stars emitting Archean-like levels of radiation, and these are primarily due to atmospheric temperature effects on particle coagulation timescales and therefore haze particle size \citep{Arney2016}. The modern Sun's planet has the peak of its reflectance spectrum at $\lambda$ $\sim$ 0.7 $\mu$m, compared to $\lambda$ $\sim$ 0.5 $\mu$m for the Archean, GJ 876, and K2V planets, and this is due to the larger size particles in the Modern Sun planet's atmosphere: the maximum radius of the Archean, GJ 876, and K2V haze particles plotted here is $\sim$0.5 $\mu$m versus $\sim$0.79 $\mu$m for the modern Sun's planet. Absorption and scattering efficiencies, Q$_{abs}$ and Q$_{scat}$, are both larger for bigger fractal particles, and Q$_{scat}$ also trends towards flatter wavelength-dependence as particle size grows (Figure \ref{fig:pod2_1}). Increased absorption (higher Q$_{abs}$) deepens the short wavelength absorption feature produced by a thick haze of larger particles. Meanwhile, the larger scattering efficiency (higher Q$_{scat}$) at longer wavelengths for larger particles increases the brightness of the planet at these wavelengths, pushing the peak of the reflectance spectrum redward. This demonstrates the need to simulate particles in coupled photochemical-climate models to capture the effects of atmospheric temperature on particle size and the resulting impacts on the planetary spectrum.
 
The impact of haze on the temperature structures of the atmospheres simulated here can also be seen in the thermal radiation spectra (middle panel of Figure \ref{fig:pod2_4}). Hazes absorb UV photons and warm the stratosphere similar to ozone on modern-day Earth. Signatures of warm stratospheres (thermal inversions) in the hazy atmospheres can be seen in the thermal emission spectra as CH$_4$ and CO$_2$ in emission (rather than absorption) near 8 and 15 $\mu$m for the Archean Sun, modern Sun, and hazy K2V spectra. As discussed in Section \ref{sec:pod2_mdwarfs}, the haze around the GJ 876 planet, however, does not produce a strong thermal inversion because its star emits less UV radiation (see also its temperature profile in Figure \ref{fig:pod2_3}). On the other hand, both of the M dwarf planets have warmer surface temperatures for the reasons discussed in Section \ref{sec:pod2_mdwarfs}, and this is apparent from the larger amounts of thermal radiation emitted by these worlds in the atmospheric window between roughly 9 and 11 $\mu$m. Ethane, a strong greenhouse gas, can be seen near 12 $\mu$m in all spectra as a photochemical consequence of the large quantities of methane in these atmospheres compared to modern day Earth.

Hazes also strongly impact transit transmission spectra. Our transit spectra (bottom panel of Figure \ref{fig:pod2_4}) include the effect of atmospheric refraction \citep{Misra2014, Misra2014a}, and this makes it impossible to the probe below 15-20 km in altitude for all model planets, including those with haze-free atmospheres. For atmospheres with haze, the minimum altitude transit observations can probe is set by the altitude where the haze becomes optically thick. Note that because all of these transit spectra sense altitudes in the stratosphere, water vapor cannot be detected. Stratospheres on traditionally habitable planets are dry; a wet stratosphere would indicate a planet undergoing a runaway greenhouse. The transit spectra of the hazy worlds exhibit a scattering slope in the visible and NIR due to a combination of haze scattering and Rayleigh scattering. The thick haze shown around the modern Sun in particular produces a relatively featureless, sloped spectrum in which absorption features from gases are obscured at visible and NIR wavelengths shorter than $\sim$2 $\mu$m. At longer IR wavelengths where the haze is relatively transparent, its impact on the transit transmission spectra is diminished, and absorption features, particularly for CO$_2$ and CH$_4$, become apparent even for the modern Sun spectrum. 

The transit spectra are sensitive to hazes that are barely detectable in reflected light due to the longer path length taken by light in transit observations. The haze around the AD Leo planet with CH$_4$/CO$_2$= 0.9 is scarcely distinguishable from a planet without haze in reflected light. However, the AD Leo haze is more apparent in the transit transmission spectrum compared to the haze-free planets. 

\begin{figure*}[!htb]
\begin{center}
\includegraphics [scale=1]{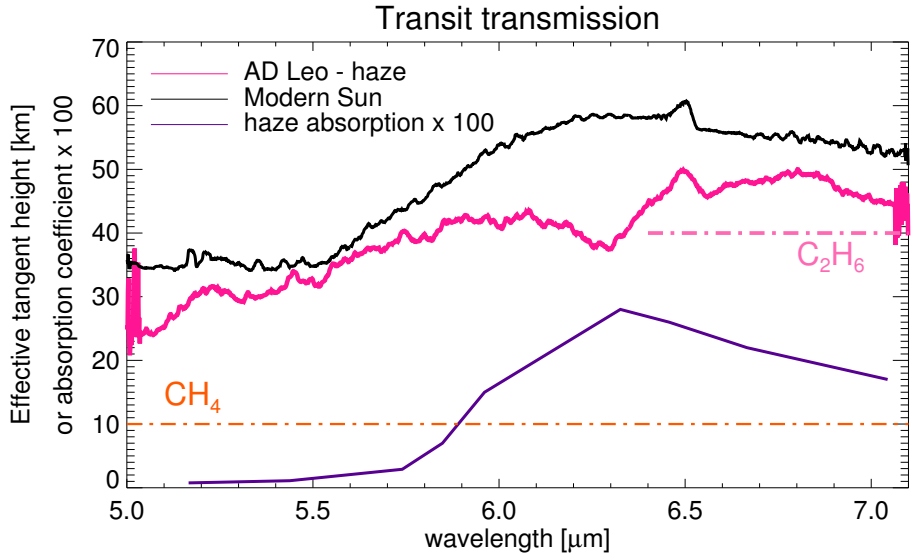}\\
\caption{A zoom in on the region around 6 $\mu$m showing the transit transmission spectra of the modern Sun planet, which has the most optically thick haze of all the planets studied here, and the ``AD Leo - haze'' spectrum, which has a very thin haze but the largest amount of C$_2$H$_6$. The purple solid line shows the haze extinction coefficient ($k$) scaled by a factor of 100 to plot on the same y-axis, and the peak in this curve corresponds to the peak in the ``modern Sun'' spectrum. Dot-dashed lines show the wavelength ranges where CH$_4$ and C$_2$H$_6$ absorb. Absorption from haze occurs near 6.3 $\mu$m in the ``modern Sun'' spectrum, and absorption from C$_2$H$_6$ prominently occurs between 6.5 and 7 $\mu$m for the ``AD Leo - haze'' spectrum.}
\label{fig:pod2_5}
\end{center}
\end{figure*}

As first discussed \citet{Arney2016}, there is an absorption feature from the haze itself near 6 $\mu$m (caused primarily by C=C and C=N stretching) that may allow remote identification of hydrocarbon hazes on exoplanets. This feature produces the increase in effective tangent height in the hazy transit transmission spectra at this wavelength. Ethane and CH$_4$ absorption overlaps with the haze's 6 $\mu$m absorption feature, but the haze feature can be distinguished by higher opacity centered around 6.3 $\mu$m. We focus on this feature in Figure \ref{fig:pod2_5}, which compares the AD Leo planet with a sparse haze to the Modern Sun planet with a thick haze. The AD Leo planet has more CH$_4$ and C$_2$H$_6$ than the modern Sun's planet. There is a peak in the haze extinction coefficient near 6.3 $\mu$m, which causes an increase in absorption for the modern Sun planet. The AD Leo planet's spectrum in this region is controlled by the behavior of the CH$_4$ and C$_2$H$_6$ absorption cross-sections because its haze is very thin.

In addition to the 6 $\mu$m feature, there is a much weaker haze absorption feature near 3 $\mu$m that is most easily seen as a small bump in the modern Sun spectrum. The weakness of the 3 $\mu$m haze feature makes it unlikely to be detectable. Both the 6 $\mu$m and 3 $\mu$m features can be seen as peaks in the haze Q$_{abs}$ curve in Figure \ref{fig:pod2_1}. These peaks appear to be general features of organic haze and are not specific to our use of the \citet{Khare1984} optical constants (see Figure 14 in \citet{Arney2016} for a comparison of haze optical constants in the literature).

\subsection{UV irradiance at the surface of hazy worlds}
Hazes are strong absorbers at UV wavelengths (Figure \ref{fig:pod2_4}) and so could potentially act as a UV shield for planetary surfaces. In particular, fractal organic hazes could have provided a UV shield for the anoxic Archean atmosphere \citep{Wolf2010, Arney2016}, especially for DNA-damaging UVC radiation ($\lambda < 0.280 \mu$m). Since the Archean likely lacked an O$_2$/O$_3$ shield, another shielding agent would have assisted the development of land-based life.   

Table \ref{tab:pod2_4} summarizes the UV flux at the surface (W/m$^2$) for UVA ($\lambda$ = 0.315 - 0.400 $\mu$m), UVB ($\lambda$ = 0.280 - 0.315 $\mu$m), and UVC ($\lambda$ $<$ 0.280 $\mu$m) radiation for each of our planets. For comparison, we also include the surface UV fluxes our model calculates for the actual modern day Earth atmosphere.  Note that the ``Archean Sun'' results presented here are not the same as the results presented for UV shielding in our earlier work \citep{Arney2016}. The haze for the ``Archean Sun'' here refers to simulations with CH$_4$/CO$_2$= 0.2 for pCO$_2$ = 0.01, and this haze is slightly thinner than the one discussed in the context of UV shielding in \citet{Arney2016}, which referred to CH$_4$/CO$_2$= 0.21 for pCO$_2$ $\sim$ 0.02. The hazy ``modern Sun'' planet has less UVA and UVB at the surface compared to the actual modern day Earth, illustrating how the broadband UV absorption by organic haze cuts down UVA and UVB far better than gases in the actual modern day atmosphere. The haze-free UV fluxes we quote here are comparable to the fluxes for similar stars found by \citet{Rugheimer2015} in a study of the UV surface environment of Earthlike planets orbiting various stellar types.

\begin{table}[]
\centering
\caption{The integrated UV fluxes (W/m$^2$) at the surface in UVA, UVB, and UVC for all of the spectra presented in our study. ``Modern Day Earth'' refers to the actual modern (haze-free) planet. All UV fluxes are presented for a solar zenith angle of 60$^\circ$. As before, all of our planets have CH$_4$/CO$_2$= 0.2 except the hazy K2V planet (CH$_4$/CO$_2$= 0.3), the hazy AD Leo planet (CH$_4$/CO$_2$= 0.9), and the modern day Earth which has the actual modern atmosphere.}
\label{tab:pod2_4}
\begin{tabular}{llll}
Star                  & UVA  & UVB     & UVC     \\
\hline
\hline
Modern Day Earth      & 29   & 0.45    & $\sim$0 \\
Modern Sun - no haze  & 29   & 5       & 1.26    \\
Modern Sun - haze     & 0.72 & 0.012   & 0.00031 \\
Archean Sun - no haze & 23   & 3.8     & 0.93    \\
Archean Sun - haze    & 8.3  & 0.76    & 0.11    \\
AD Leo - no haze      & 0.41 & 0.041   & 0.043   \\
AD Leo - haze         & 0.37 & 0.035   & 0.034   \\
GJ 876 - no haze      & 0.53 & 0.0051  & 0.0031  \\
GJ 876 - haze         & 0.18 & 0.00079 & 0.00018 \\
K2V - no haze         & 13   & 2.1     & 0.29    \\
K2V - haze            & 3.5  & 0.27    & 0.02    \\
F2V - no haze         & 38   & 8.6     & 4.6    
\end{tabular}
\end{table}

The surface UVC fluxes of the ``Modern Sun - haze'' planet and the ``GJ 876 - haze'' planet are higher than we currently experience on Earth but should be easily tolerated by \textit{Chloroflexus aurantiacus}, an anoxygenic phototroph that has been studied  an analog for Archean photosynthetic organisms \citep{Pierson1992}. \textit{Chloroflexus aurantiacus} was shown in \citet{Pierson1992} to exhibit moderate growth under UVC fluxes comparable to or lower than the fluxes calculated here for every  star in Table \ref{tab:pod2_4} except the F2V star, the modern Sun with no haze, and the Archean Sun with no haze. Of course, life can also take refuge from UV radiation under other types of chemical or physical UV shields (e.g., within a liquid water column) so even these higher UV fluxes do not necessarily prohibit life \citep{Cockell1998}. Still, UV shielding is an important consideration for planetary habitability, so despite their cooling effects, UV-blocking hazes like the ones studied here may actually enhance planetary habitability. 

Our analysis does not consider M dwarf flaring events, which can increase the UV irradiance by orders of magnitude \citep{Segura2010}. Since we have shown that stars with very high UV flux -- particularly high FUV fluxes --  do not form hazes as readily or at all compared to stars with lower FUV fluxes, frequent flaring events are expected to have a deleterious effect on a haze layer, although we have not examined the effects of time-dependent flares here. 

\subsection{Detectability of organic haze}
Organic haze's strong absorption features provide an indirect way to sense atmospheres rich in CH$_4$ even if the CH$_4$ absorption features themselves are not distinguishable. Because attempts to characterize exoplanets have been frustrated by the presence of atmospheric aerosols \citep[e.g.,][]{Kreidberg2014}, haze is typically considered to obscure planetary characteristics. However, for the organic hazes presented here, gaseous absorption features can still be seen for $\lambda$ $>$ 0.5 $\mu$m in reflected light and for $\lambda$ $>$ 1 $\mu$m in transit transmission even in the hazy spectra. 

Although they may obscure aspects of the planetary environment, organic hazes have the potential to unveil interesting ongoing planetary processes. The presence of an organic haze implies an active source of methane, particularly in high CO$_2$ atmospheres like Archean Earth, which requires a CH$_4$/CO$_2$level $>$ 0.1  -- and therefore a substantial CH$_4$ flux on the order of $\sim1\times10^{11}$ molecules/cm$^2$/s before haze formation occurs. This methane flux is comparable to the methane production rate by biology on modern Earth \citep{Kharecha2005}, so hazes in atmospheres with Archean-like CO$_2$ levels could signal possible biological activity.  

Because methane can be produced by a variety of biological and non-biological means, there is no reason to expect organic-rich planets in the habitable zone to be rare. We should therefore be prepared for the detection of hazy habitable planets orbiting G, K, and M dwarfs. In this section, we discuss the detectability of organic haze around M dwarfs with JWST, and around G and K dwarfs with a future large 10-m direct imaging telescope.

\subsubsection{Simulated JWST observations}\label{sec:pod2_jwst}
M dwarf planet hosts will be important targets for transit transmission observations by the James Webb Space Telescope (JWST) because the ratio of the planet's size relative to the star's size is largest for M dwarfs. Thus, their transit transmission signals are larger compared to equal-radius planets orbiting higher mass stars. Habitable zone planets also orbit closer to M dwarf stars, so their transits occur more frequently than for planets orbiting higher mass stars. 

\begin{figure*}[!htb]
\begin{center}
\includegraphics [scale=0.8]{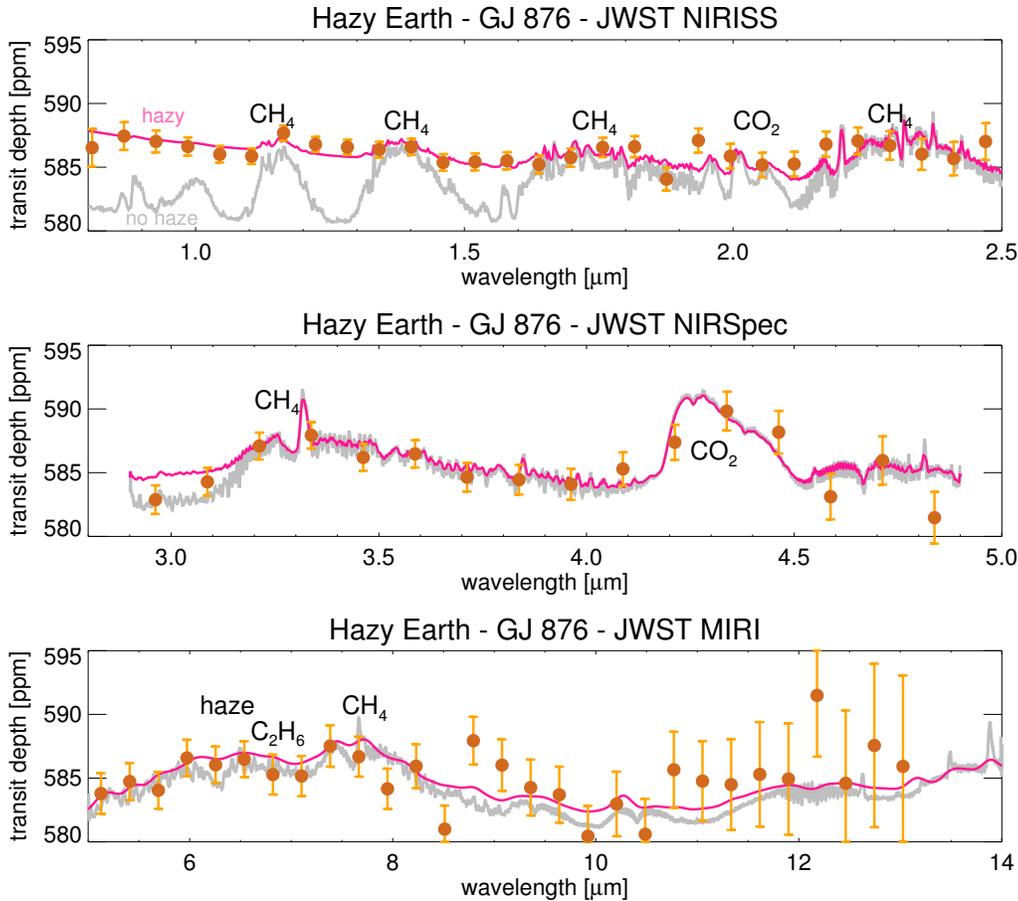}\\
\caption{Shown are simulated transit spectra as seen by JWST for our hazy GJ 876 planet. The top panel simulates the NIRISS instrument, the middle panel simulates the NIRSpec instrument, and the bottom panel simulates the MIRI instrument. The pink spectrum shows the full spectrum prior to being fed into the simulator. The orange points with error bars are the spectrum as seen by JWST over 10 transits (65 hours of integration time). The gray spectrum shows a haze-free planet for comparison.}
\label{fig:pod2_6}
\end{center}
\end{figure*}

Figure \ref{fig:pod2_6} shows the results of our simulated observations over 65 hours of integration time (10 transits) per instrument for a planet orbiting GJ 876. The pink line shows the simulated spectrum, and the orange points with error bars denote the simulated JWST observations. The gray line shows the planet without haze, which is included for comparison. The error bars are calculated assuming photon limited noise, which is the same assumption made in \citet{Schwieterman2016}. The large error bars at wavelengths longer than 8 $\mu$m is due to spectral noise caused by the dim stellar blackbody at these wavelengths.

We calculate the continuum level for the JWST observations by fitting a polynomial to continuum regions. To determine the detectability of spectral features, we determine the continuum level around absorption features, then subtract off that continuum. We calculate signal-to-noise (SNR) of absorption features in our simulations by binning across the features and comparing the signal to the noise level. The noise level is computed from the error on the binned absorption features and the error on the continuum estimate added in quadrature. 

CH$_4$ and CO$_2$ can be detected across the NIRISS and NIRSpec bands. The CH$_4$ feature near 1.7 $\mu$m has SNR = 3.0, and CO$_2$ at 2 $\mu$m has SNR = 2.8. Detections of shorter wavelength features are $< 1\sigma$. However, the broader and stronger CH$_4$ feature near 2.3 $\mu$m is detectable at SNR = 6.4. For the NIRSpec absorption features, the opacity of the haze is negligible, and we measure SNR = 6.4 for the CH$_4$ feature near 3.3 $\mu$m and SNR = 5.9 for the CO$_2$ feature near 4.3 $\mu$m. In principle, therefore, it will be possible to measure CH$_4$ and CO$_2$ abundances and the CH$_4$/CO$_2$ ratio in transit transmission for the atmospheric altitudes probed by transit observations.  

The haze is more transparent at the long wavelength end of the of the NIRISS bandpass and in NIRSpec. The abundances of CH$_4$ and CO$_2$ inferred from longer wavelengths where the haze is less opaque will be larger than the gas abundances inferred from the absorption features at shorter NIRISS wavelengths that are truncated by the haze. Based on the size of the error bars in the continuum regions around absorption bands, the absorption features in the hazy spectrum for $\lambda$ $<$ 2.5 $\mu$m are about 2-10$\sigma$ shallower than they would be in a spectrum without haze. The inconsistency between the retrieved gas abundances at longer wavelengths compared to abundances retrieved at shorter wavelengths would suggest the presence of a haze whose opacity increases towards shorter wavelengths. 

Detection of spectral features in MIRI is in principle more challenging than NIRISS and NIRSPec because of the decline of the stellar blackbody at these longer wavelengths. However, if we define the MIRI continuum from between the first point in the MIRI bandpass and the points between 8-10 $\mu$m, we can measure SNR = 5.1 for the set of absorption features between 6-8 $\mu$m, which includes the haze absorption feature. Higher signal-to-noise would be needed to distinguish the haze from other absorbers in this region, but the presence of haze can be inferred separately from the NIRISS and NIRSpec observations of the CH$_4$/CO$_2$ ratio and the depths of the NIRISS absorption features compared to the expected haze-free level.

The instrument models above do not include any contributions from systematic noise.  Systematic noise sources come, in large part, from instrumentation and detectors, and will not be fully characterized until after launch. They will also tend to decrease in time as instrument and detector models are improved and new observing techniques are developed. The simulated observations presented here indicate that achieving a combined noise (random plus systematic) at the level of several ppm will be essential for characterizing hazy exo-Earths. Such precision may be possible if the systematic noise sources are characterized to a level well below the random noise. However, if the JWST systematic noise represents a floor at the $>$ 10 ppm level, as proposed by \citet{Greene2016}, then characterizing the hazy exo-Earths presented here becomes extremely difficult, and the error bars on our simulated JWST measurements will become much larger.

\subsubsection{Simulated direct imaging observations}
Unlike for JWST observations, M dwarf planets hosts are generally poor targets for direct imaging surveys because it will likely not be possible to angularly separate their habitable planets from their host stars except for the closest M dwarfs (e.g. Proxima Centauri b \citep{Anglada2016}). The inner working angle (IWA), which defines the smallest angular separation between a planet and its host star at which the planet can be detected, scales with $\sim$N$\lambda$/D where D is the telescope diameter and N is a small-valued constant of order 10$^0$. F dwarf habitable planets, which will naturally orbit farther from their stars than planets orbiting cooler hosts \citep{Kopparapu2013}, are most likely to be observable outside the IWA for the star types we simulate here, but we have shown that Archean-like worlds orbiting F dwarfs are less likely to have organic hazes. Note F dwarfs are less numerous than lower mass stars, so the distance to any one of them is likely to be larger than for G, K, and M dwarfs, and so the planet-star angular separation may still pose a problem. G and K dwarf planets, on the other hand, may have organic haze, and such stars will be important targets for future direct imaging missions \citep{Stark2014}.

\begin{figure*}[!htb]
\begin{center}
\includegraphics [scale=0.8]{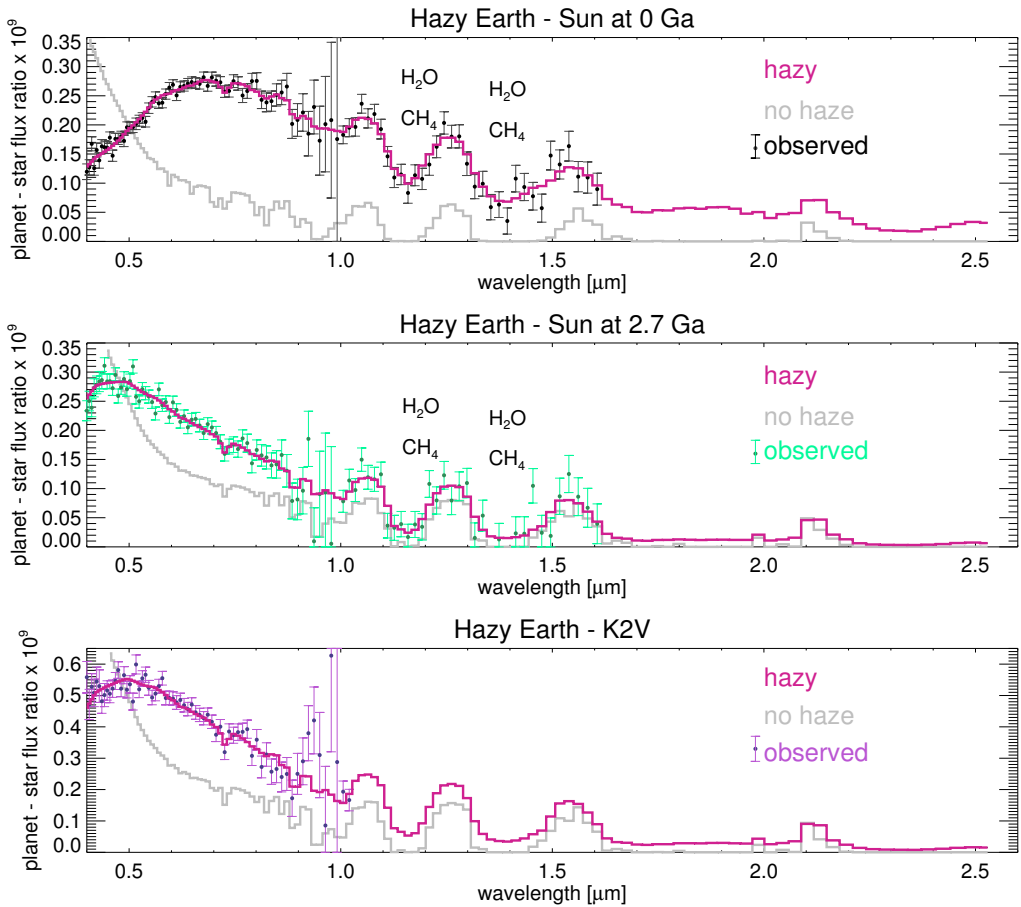}\\
\caption{These are simulated reflectance spectra as seen by a 10-m LUVIOR-type telescope over 200 hours of integration time in each spectral band for planets at 10 pc. The colored lines show the hazy spectra without noise added, and the gray spectra are the corresponding haze-free planets for comparison. Points with error bars simulate the hazy spectrum seen by the telescope with realistic noise sources. }
\label{fig:pod2_7}
\end{center}
\end{figure*}

We tested what a hazy Archean Earth analog orbiting the modern Sun, the Archean Sun, and the K2V dwarf would look like to a future 10-m LUVIOR-type space telescope \citep{Postman2010, Bolcar2015, Dalcanton2015} using the coronagraph instrument noise model described in \citet{Robinson2016}. The star-planet systems are assumed to be located at a distance of 10 parsecs. The results of these simulations are presented in Figure \ref{fig:pod2_7}. A spectrum with the haze removed (gray line) is presented alongside the hazy spectra (pink line) for comparison in all three cases. The ``observed'' spectra are simulated assuming 200 hours (roughly 1 week) of integration time per coronagraphic bandpass (which may not span the entire wavelength region of interest) for a planet at quadrature. If the planets were at a distance of 3 pc instead of 10 pc, the integration time needed to achieve the same signal-to-noise decreases by about an order of magnitude; we use 10 pc here to be conservative and to be able to consider the challenges of detecting distant planets. The spectral resolution (R = $\lambda$/$\Delta \lambda$) is 70, and the telescope wavelength range is 0.4 - 3 $\mu$m. We chose an outer working angle of  20$\lambda$/D, and an inner working angle of 3$\lambda$/D. For a 10 m mirror, the inner working angle  limits the longest wavelengths that can be observed in all three cases: the planets orbiting the G2V star cut off near 1.5 $\mu$m, and the planet orbiting the K2V star cuts off near 1 $\mu$m. Visible and NIR wavelength ranges are observed by separate detectors as described in \citet{Robinson2016}. It is assumed that the telescope system will be cooled to a sufficiently low temperature to minimize detector thermal noise (T $<$ 80 K) that would otherwise contribute to spectral noise in the NIR. Thermal noise should not contribute appreciably to wavelengths $<$ 1.6 $\mu$m, so our assumption of a cold telescope should not strongly impact the results shown here.

 A throughput of 5\% is assumed from the work of \citet{Robinson2016}, although LUVOIR may have a higher throughput closer to 20\% (A. Roberge, personal communication). For this reason, the simulations presented here may be considered conservative.

To test the detectability of spectral features, we followed a similar procedure to our JWST analysis. We fit a polynomial to the continuum regions near features of interest, then removed the continuum using the procedure described above. We then binned across the absorption features to determine their SNR compared to the continuum level.

We find that, for all three stars, the haze absorption feature at UV-blue wavelengths is easily detected against the expected extrapolated continuum level. The modern Sun haze absorption feature has an extremely robust detection of SNR = 70, the Archean Sun haze has SNR = 12, and the K2V haze has SNR = 27. Our extrapolated continuum only extends the polynomial fit to longer wavelengths and does not include the expected Rayleigh scattering, so these detections are under-estimates compared to a model that includes Rayleigh scattering. An absorption feature from overlapping CH$_4$ and H$_2$O can be seen near 0.72 $\mu$m for all three planets. It can be detected at SNR = 3.1 for the modern Sun, SNR = 2.6 for the Archean Sun, and SNR = 3.3 for the K star.

Features can be seen in the modern and Archean Sun spectra near 1.15 and 1.4 $\mu$m that are caused by overlapping H$_2$O and CH$_4$ absorption bands. The 1.15 and 1.4 $\mu$m features are detected at SNR = 16 and SNR = 11 for the modern Sun, and at SNR = 11 and SNR = 9 for the Archean Sun, respectively. At 10 pc, these features are cut off by the K star inner working angle and cannot be observed. Because of the $\sim$1 $\mu$m IWA cutoff for a K star planet at 10 pc, the haze absorption band is the strongest feature that can be seen, providing indirect evidence of methane. 

The large error bars around 1 $\mu$m are caused by falloff in both CCD and InGaAs detector quantum efficiencies assumed by the simulator, making it difficult to detect spectral features here. In the NIR, considering the frequent overlap between CH$_4$ and H$_2$O bands, to detect and quantify the abundance of water even for the G star spectra, good sensitivity to the clean, methane-free water bands near 0.82 $\mu$m and especially the stronger band near 0.94 $\mu$m (Figure \ref{fig:pod2_4}) is crucial. Because water could not be cleanly detected in these spectra with the assumed detector technology, retrievals of gas abundances would likely exhibit degeneracies in the retrieved amounts of H$_2$O and CH$_4$.

Note that organic haze itself could be an indirect sign of water on Earthlike planets. Because sources of methane on an Earthlike planet are likely to involve water (either through biological production or serpentinization, the dominant abiotic methane source on Earth, and one which requires water), the indirect detection of methane on a terrestrial planet could also be argued to suggest the presence of water, particularly in an atmosphere with CO$_2$ that necessitates a vigorous CH$_4$ flux to produce haze. 

CO$_2$ is important to detect for this reason and others. For instance, measuring CO$_2$ abundance can constrain the redox state of the atmosphere, and its presence in a planetary atmosphere can help determine whether a planet is in fact terrestrial in the absence of other data for mass or planetary radius. Unfortunately, CO$_2$ cannot be detected in reflected light for any planets shown here. Although there is a CO$_2$ feature near 1.6 $\mu$m, it is not detectable at the spectral resolution and noise level we simulate. The strongest CO$_2$ band shortward of 5 $\mu$m is near 4.3 $\mu$m. However, this band would be difficult to measure in direct imaging. We assume a cold telescope in the simulations here, but thermal radiation from a telescope that is not cryogenically cooled would be very significant for wavelengths longer than about 1.8 $\mu$m. In addition, 4.3 $\mu$m may not be accessible with the telescope's inner working angle (as in the examples shown here). Assuming IWA = 3$\lambda$/D, a telescope would have to be about 27 m in diameter to reach 4.3 $\mu$m for a planet that is 1 AU from its star at a distance of 10 pc. On the other hand, a 10 m mirror would be sufficient to reach  4.3 $\mu$m for a target at 3 pc.

Longer wavelengths such 4.3 $\mu$m may be more easily observable with LUVOIR for targets whose geometry allows for transit transmission observations. It may also be possible to observe exoplanets such as the ones simulated here with the next generation of ground-based observatories with larger mirror diameters than LUVOIR using adaptive optics and advanced coronography. These observatories will have to contend with spectral contamination from Earth's atmosphere, but \citet{Snellen2013} has suggested that the Doppler shift of the planet could be used to disentangle its spectrum from Earth's atmosphere.

\section{Discussion}
We have found that organic haze should be detectable on nearby Archean-analog exoplanets with future space-based telescopes. Here, we discuss some of the limitations of our model's haze formation scheme. We also discuss implications of haze's spectral features with respect to cooling of planetary surface environments, and we compare the haze's UV-blue absorption feature to other UV-blue absorbers. Lastly, we discuss how oxygen spectral features are not detectable in our planetary atmospheres despite oxygen production around some stars that frustrates haze formation.

\subsection{Limitations of haze formation in our photochemical scheme}
Our results suggest that G dwarfs, K dwarfs, and some M dwarfs are more likely to generate hydrocarbon hazes in Earthlike atmospheres compared to F dwarfs and stars with frequent flare events such as AD Leo. To generate hazes, stars need sufficient UV flux to drive the relevant photochemistry through reactions such as CH$_4$ + h$\nu$($\lambda$ $<$ 150 nm) $\rightarrow$ CH$_3$ + H, but too much FUV flux generates oxygen radicals through reactions such as CO$_2$ + h$\nu$ ($\lambda$ $<$ 200 nm) $\rightarrow$ CO + O and  CO$_2$ ($\lambda$ $<$ 200 nm) + h$\nu$ $\rightarrow$ CO + O$^{1}$D that halt the haze formation process by oxidizing hydrocarbon photochemical products.

However, our model assumes a mechanism proposed for the formation of Titan's haze \citep{Allen1980, Yung1984} such that haze formation occurs through formation of acetylene (C$_2$H$_2$) and its further polymerization to higher order hydrocarbons. In reality, this scheme is likely overly simplistic. For example, measurements of Titan's hazes by the Cassini spacecraft have discovered nitrile chains and nitrogen-bearing polycyclic aromatic hydrocarbons (PAHs) that suggest nitrogen-bearing compounds may be important to haze formation on Titan \citep{Waite2007, Lopez-Puertas2013}.

Titan's atmosphere is extremely reducing, but Archean Earth's atmosphere was probably less so, containing non-negligible amounts of CO$_2$ \citep{Kasting1993, Driese2011}. Interestingly, laboratory experiments have suggested that the presence of oxygen may not be as harmful to haze production as our haze-formation scheme suggests here. For example, \citet{Trainer2006} showed that haze formation in a CH$_4$/N$_2$ mixture containing CO$_2$ was more efficient than in a CH$_4$/N$_2$-only mixture because the oxygen atoms produced by CO$_2$ photolysis were incorporated into the haze molecules. Furthermore, \citet{DeWitt2009} showed the existence of carbonyl and carboxyl groups in aerosol analogs with C/O = 0.1. \citet{Horst2014} showed that CO, also, can benefit aerosol formation and be a source of oxygen incorporation into aerosol molecules. Recently, \citet{Hicks2016} showed oxygen from CO$_2$ incorporated into haze molecules can comprise 10\% of the mass of Archean haze particles.

These complexities suggest that an updated study incorporating these mechanisms into our photochemical model will be necessary to determine their impact on haze formation for the planets simulated here. We may find that haze formation is enhanced relative to our findings for planets orbiting stars with efficient oxygen-production, and the hazes that form in more oxygen-rich atmospheres may differ in composition and spectral properties compared to those in more oxygen-poor atmospheres. Updates to our photochemical model including incorporation of laboratory studies we are involved with will allow us to examine these issues are part of our ongoing work and future work.

\subsection{Haze-induced cooling of planetary surfaces}\label{sec:pod2_cooling}
Since haze can cool a planetary climate, there may be a ``hazy habitable zone'' (HHZ) inner edge closer to the star than the traditional habitable zone boundaries \citep{Kasting1993, Kopparapu2013} for planets with organic-rich atmospheres. However, the results presented here indicate that the inner edge of the HHZ will not be relevant to certain types of stars such as F and M dwarfs.  As we have seen, F2V planets with atmospheres containing CO$_2$ and H$_2$O do not generate this haze even at high CH$_4$/CO$_2$ ratios due to the buildup of haze-destroying oxygen-containing species. Some M dwarf planets are able to generate haze for the types of atmospheres considered here, but its cooling effects would be small because the M dwarf spectral output is in a wavelength range where these hazes are relatively transparent.  

The outer edge of the habitable zone (OHZ) may also be affected by organic hazes. The OHZ is traditionally defined as the distance where CO$_2$ greenhouse warming is balanced by Rayleigh scattering from additional CO$_2$. In principle, the warming potential for organic-rich planets at the outer edge of the habitable zone would be limited by the formation of haze and its attendant antigreenhouse effect. This process could define the ``maximum greenhouse effect'' for organic-rich worlds orbiting G and K stars. However, all of this is subject to the caveat that habitable planets near the OHZ may not be able to generate organic hazes in the first place. If the maximum CO$_2$ greenhouse limit allows for several bars of CO$_2$, implausibly large CH$_4$ fluxes may be required to achieve a high enough CH$_4$/CO$_2$ ratio to create a haze in such atmospheres.

\subsection{A comparison to other aerosols and UV-absorbers}

\begin{figure*}[!htb]
\begin{center}
\includegraphics [scale=1]{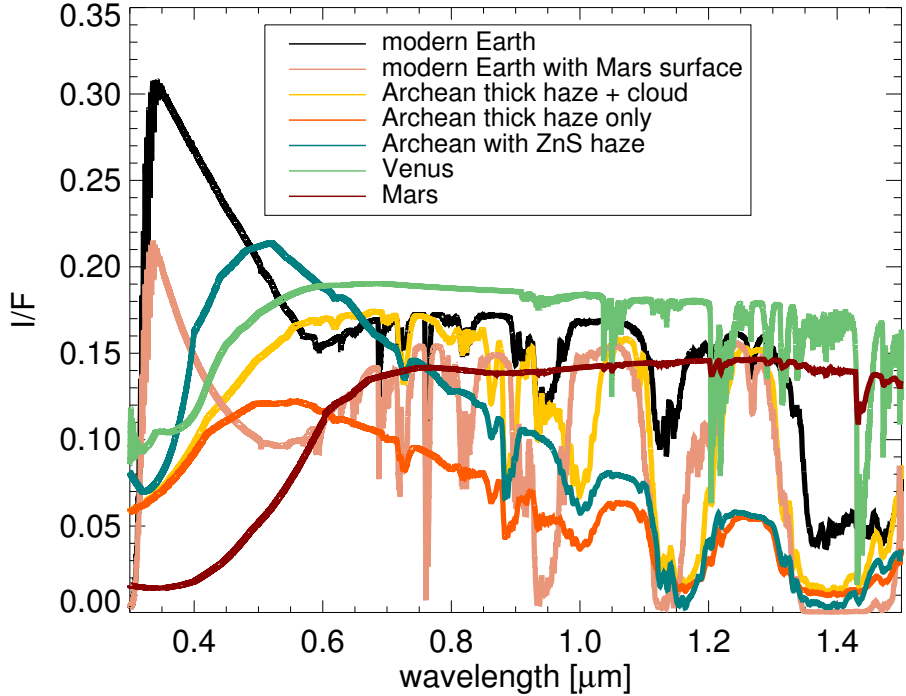}\\
\caption{This shows a comparison of several types of spectra with broad short wavelength absorption features to compare haze's short wavelength absorption feature to similar features produced by different types of absorbers. These spectra have been scaled in the y-dimension to plot together on the same axis.  }
\label{fig:pod2_8}
\end{center}
\end{figure*}

The haze's broadband UV and blue wavelength absorption feature is prominent and distinctive in reflected light, but to ensure accurate interpretation of this feature, it is important to explore similar UV absorbers that might mimic this feature in a planet's spectrum. We compare this haze's spectral signature with other short wavelength absorbers in Figure \ref{fig:pod2_8}, which plots hazy Archean Earth alongside modern Earth with clouds, Venus, Mars, modern Earth with a Mars-like surface, and Earth with a ZnS haze. The ZnS spectrum is not intended to be physically realistic and is provided simply to show the absorptive effects of ZnS particles, which have strong UV absorption similar to organic haze. We also show hazy Archean Earth with water clouds constructed using a weighted average of 50\% haze-only, 25\% haze and cirrus cloud, and 25\% haze and strato cumulus cloud \citep[and see also our discussion of water clouds in hazy Archean spectra in \citet{Arney2016}]{Robinson2011}.

In the modern Earth atmosphere, the ozone Chappuis band is a broad feature centered near 0.5-0.7 $\mu$m, but its absorption does not continue farther into the UV the way haze does, so the Rayleigh scattering slope becomes prominent for $\lambda$ $<$ 0.5 $\mu$m, distinguishing this spectrum from a hazy one.  In addition, hydrocarbon hazes are unlikely to be present in atmospheres with spectrally apparent ozone (Section \ref{pod2_oxy}).

Mars is red because iron oxide absorbs strongly at blue wavelengths.  A spectrum of the Earth with the wavelength-dependent surface albedo of Mars shows what Earth could look like with a surface rich in iron oxide. On an Earthlike planet with a 1 bar atmosphere, the blue-absorbing iron oxide feature is unlikely to be mistaken for hydrocarbon haze due to increased reflectivity for $\lambda$ $<$ 0.5 $\mu$m due to Rayleigh scattering. It is also important to note that at low spectral resolution, iron oxide could also be mistaken for ozone absorption. However, on Mars itself, Rayleigh scattering is non-apparent, so the strong iron oxide absorption feature could mimic haze. Mars's iron oxide feature could be distinguished from haze by the absence of CH$_4$ features in the spectrum.

Earth with a thick haze of ZnS particles is the closest mimic we found to our cloud-free spectrum of Archean Earth with a hydrocarbon haze. However, ZnS is not a realistic aerosol candidate for Earthlike atmospheres because it condenses at temperatures close to 1000 K \citep{Morley2012}. \citet{Charnay2015} and \citet{Morley2015} show what a ZnS haze would look like in a more realistic atmosphere for GJ 1214b. Therefore, constraints on surface temperature or semi-major axis could eliminate this as a potential source of UV absorption.

Venus's broad UV absorption caused by its unknown UV absorber \citep{Markiewicz2014} and SO$_2$ can also mimic the UV absorption of organic haze.  However, the Venus spectrum lacks CH$_4$ features and strong water features that would indicate habitability. Venus' oxidizing atmosphere is very different from Archean Earth. 

Organic haze's blue and UV wavelength absorption feature together with observations of methane bands would strongly imply the existence of haze in an atmosphere. The UV absorbers we compared to here can be distinguished from organic haze through the appearance the Rayleigh scattering slope, the lack of CH$_4$ features, or (in the case of ZnS) are extremely unlikely for an Earthlike atmosphere. 

\subsection{Detectability of Photochemical Oxygen}\label{pod2_oxy}
Although the F2V planet produces a significant amount of oxygen radicals compared to our other stars, the absolute level of oxygen in its atmosphere is not large enough to be detectable. The column densities of O$_2$ and O$_3$ in the F2V atmosphere are $9.0 \times 10^{18}$ molecules/cm$^2$ and $2.61 \times 10^{14}$ molecules/cm$^2$, respectively. The column density of methane is $4.27 \times 10^{22}$ molecules/cm$^2$. As discussed in \citet{Domagal-Goldman2014}, it is difficult to accumulate abiotic oxygen at detectable levels in atmospheres rich in organics because reactions with reduced gases are major oxygen sinks. \citet{Domagal-Goldman2014} show that at lower CH$_4$ column densities than the ones we simulate here (e.g. $\sim$10$^{16}$ molecules/cm$^2$), O$_2$ and O$_3$ can reach column densities of $10^{18}-10^{21}$ molecules/cm$^{2}$ and $10^{16}-10^{18}$ molecules/cm$^2$, respectively, and O$_3$ can produce significant spectral signatures. Another study, \citet{Harman2015}, showed that O$_2$ and O$_3$ from CO$_2$ photolysis can produce spectral signatures in atmospheres with low CH$_4$ mixing ratios different from those simulated here. However, the \citet{Harman2015} model produces similar O$_3$ and O$_2$ column depths when using similar assumptions to ours for the CH$_4$ and CO$_2$ mixing ratios and broadly reproduces the trends seen here between different stellar types (C. Harman, personal communication).
 \\
\section{Conclusions}
Hazy earthlike planets may be common, so the conditions that form haze, and the haze's climatic and spectral effects are important to understand. We have shown the likelihood that a planet will form organic hazes varies strongly with host star spectral type. Stars with very high FUV fluxes (e.g., F stars) are seem unlikely to form organic haze due to the buildup of oxygen species that destroy hydrocarbons. Future work with more complete photochemistry that includes oxygen incorporation into haze molecules will allow us to test this conclusion with a more complex photochemical scheme. For planets with haze, antigreenhouse cooling is important to G and K dwarf planets, but because M dwarfs emit the bulk of their radiation at wavelengths where these hazes are relatively transparent, haze-induced cooling for M dwarf planets is insignificant. Organic haze produces distinctive absorption features, including an absorption feature near 6.3 $\mu$m that may be detectable with JWST. A strong UV and blue wavelength absorption feature may provide a UV shield for surface biospheres and could be detected with a proposed large direct imaging space-based telescope like LUVOIR.

Hydrocarbon haze may also be a more detectable indication of high CH$_4$ abundances in terrestrial planetary atmospheres than the CH$_4$ itself. Finding an organic haze in the atmosphere of a planet with Archean-like CO$_2$ levels would be indicative of highly interesting processes that imply ongoing geological and/or biological activity. Although haze is often considered to be a feature that conceals certain atmospheric features and surface processes, in this case the haze itself can indicate a geologically active planet -- and therefore a potentially habitable one -- and possibly even reveal the presence of life. 

\acknowledgments
We are grateful to our anonymous reviewer for their thorough and extremely useful comments that improved our manuscript. This work was performed as part of the NASA Astrobiology Institute's Virtual Planetary Laboratory, supported by the National Aeronautics and Space Administration through the NASA Astrobiology Institute under solicitation NNH12ZDA002C and Cooperative Agreement Number NNA13AA93A. E.T. Wolf acknowledges NASA Planetary Atmospheres Program award NNH13ZDA001N-PATM and NASA Exobiology Program award NNX10AR17G for financial support. Simulations were facilitated through the use of the Hyak supercomputer system at the University of Washington eScience Institute. We thank Dave Crisp, Thomas Gautier, Sonny Harman, Jacob Lustig-Yaeger, Aki Roberge, and the whole VPL team for useful conversations and advice on this project.  Spectra shown in this work will be archived at the Virtual Planetary Laboratory online spectral database.

\textit{Software}: Atmos \citep{Arney2016}, SMART \citep{Meadows1996, Crisp1997}, JWST model, \citep{Deming2009}, Coronagraph Noise Model \citep{Robinson2016}, IDL, Python.

\bibliography{apjmnemonic,refs}
\bibliographystyle{apj}

\end{document}